\documentclass[prb,twocolumn,aps,amsmath,amssymb,amsfonts,superscriptaddress,preprintnumbers,floatfix]{revtex4}
\usepackage[german,american,english]{babel}
\usepackage{graphicx}
\usepackage{graphics}
\usepackage{dcolumn}
\usepackage{bm}
\usepackage{latexsym}
\usepackage{amssymb}
\usepackage{amsmath}
\usepackage{amsfonts}
\usepackage{layout}
\usepackage{verbatim}
\usepackage{epsfig}

\newcommand {\ket} [1] {| #1 \rangle}
\newcommand {\bkt} [1] {\langle #1 \rangle}

\newcommand {\tbkt} [3] {\langle #1 | #2 | #3 \rangle}

 \newcommand {\beq}{\begin{equation}}
\newcommand {\eeq}{\end{equation}}
\begin{document}
\title{Interface roughness, valley-orbit coupling and valley manipulation in quantum dots}
\author{Dimitrie Culcer}
\affiliation{Condensed Matter Theory Center, Department of Physics, University of Maryland, College Park MD20742-4111}
\affiliation{Joint Quantum Institute, Department of Physics, University of Maryland, College Park MD20742-4111}
\author{Xuedong Hu}
\affiliation{Condensed Matter Theory Center, Department of Physics, University of Maryland, College Park MD20742-4111}
\affiliation{Joint Quantum Institute, Department of Physics, University of Maryland, College Park MD20742-4111}
\affiliation{Department of Physics, University at Buffalo, SUNY, Buffalo, NY 14260-1500}
\author{S.~Das Sarma}
\affiliation{Condensed Matter Theory Center, Department of Physics, University of Maryland, College Park MD20742-4111}
\affiliation{Joint Quantum Institute, Department of Physics, University of Maryland, College Park MD20742-4111}
\begin{abstract}
We present a systematic study of interface roughness and its effect on coherent dynamical processes in quantum dots. The potential due to a sharp, flat interface lifts the degeneracy of the lowest energy valleys and yields a set of valley eigenstates. Interface roughness is characterized by fluctuations in the location of the interface and in the magnitude of the potential step. Variations in the position of the interface, which are expected to occur on the length scale of the lattice constant, reduce the magnitude of the valley-orbit coupling. Variations in the size of the interface potential step alter the magnitude of the valley-orbit coupling and induce transitions between different valley eigenstates in dynamics involving two (or more) dots. Such transitions can be studied experimentally by manipulating the bias between two dots and can be detected by charge sensing. However, if the random variable characterizing the position of the interface is correlated over distances of the order of a quantum dot, which is unlikely but possible, the \textit{phase} of the valley-orbit coupling may be different in adjacent dots. In this case tunneling between like and opposite valley eigenstates is in effect a random variable and cannot be controlled. We suggest a resonant tunneling experiment that can identify the matrix elements for tunneling between like and opposite valley eigenstates. 
\end{abstract}
\date{\today} 
\maketitle

\section{Introduction}

Of all semiconductors investigated in quantum computation (QC) only C, Si, and Ge in group IV have isotopes with zero nuclear spin, allowing long coherence times, with Si long known as an outstanding candidate. \cite{Feher_PR59} QC architectures are based on donor electron or nuclear spins in Si:P,\cite{Kane_Nature98, Vrijen_PRA00} single electron spins in gate-defined quantum dots (QDs) in Si/SiGe\cite{Friesen_PRB04, Hayes_09} or Si/SiO$_2$,\cite{Liu_PRB08, Nordberg_MOS_PRB09, Ferrus_09} and SiGe nanowires, \cite{MarcusGroup_NatureNano07} and singlet-triplet qubits. \cite{Petta_Science05} Experimentally, significant progress has been made recently in Si:P donor-based devices\cite{Andresen_NanoLett07, Mitic_Nanotech08, Kuljanishvili_NP08, Lansbergen_NP08, Fuhrer_NanoLett09, Stegner_NP06, Angus_NL07,Mottonen_PRB10, Lim_APL09} and gate-defined Si quantum dots. \cite{Rokhinson_PRL01, Goswami_NP07, Shaji_NP08, Liu_PRB08, Nordberg_MOS_PRB09, Lim_APL09, Nordberg_APL09} Graphene has also been proposed for QC,\cite{Trauzettel_NP07} and observation of the Coulomb blockade effect in a graphene quantum dot has been reported.\cite{Guo_CB_09}  Carbon nanotube quantum dots offer another physically interesting possibility, with promising experimental advances. \cite{Churchill_PRL09} The theory of QDs in group IV materials has also taken off in recent years.\cite{Friesen_PRB10, Qiuzi_PRB10, Culcer_PRB09, Culcer_10, Palyi_PRB09, Palyi_10}

The conduction band of group IV semiconductors consists of a series of equivalent valleys: bulk Si has six, Ge has four, whereas in C bulk diamond has six valleys while graphite, graphene and nanotubes have two. This multiplicity is known to affect spin quantum computing schemes in Si:P donors, \cite{Koiller_PRL01, Wellard_PRB05} Si QDs, \cite{Friesen_PRB10, Culcer_PRB09, Culcer_10} and carbon nanotube QDs. \cite{Palyi_PRB09, Palyi_10} Confinement partially lifts the valley degeneracy, and it is often necessary to consider only two valleys. In Si (001) the $\hat{\bm z}$-confinement lowers the energy of the $z$-valleys, perpendicular to the interface, by several tens of meV relative to those in the plane of the interface. The interface potential produces a valley-orbit coupling $\Delta$ between the two $\hat{\bm z}$-valleys, \cite{Ando_PRB79} which is sample-dependent. \cite{Friesen_PRB10} Measurements of $\Delta$ have been performed on quantum Hall systems \cite{Lai_PRB06, Nordberg_APL09} and at low magnetic fields.\cite{Goswami_NP07} A large valley splitting was recently reported in Si/SiO$_2$ \cite{Takashina_PRL06} and work is underway to confirm this claim. 

The valley-orbit coupling has been studied theoretically within the effective mass approximation (EMA) by many groups. \cite{Sham_PRB79, Ohkawa_JPSJ77a, Ohkawa_JPSJ77b, Ohkawa_JPSJ77c, Boykin_APL04, Friesen_PRB04, Nestoklon_PRB06, Saraiva_PRB09, Friesen_PRB10} Ref.\  \onlinecite{Sham_PRB79} determined $\Delta$ using a hard-wall boundary, requiring the wave function to vanish at the interface, whereas in Refs.\ \onlinecite{Nestoklon_PRB06, Saraiva_PRB09, Friesen_PRB10} the wave function was allowed to leak into the insulator. Ref.\ \onlinecite{Friesen_PRB10} highlighted the importance of interface miscuts in reducing $\Delta$. Beyond the EMA, Ref.\ \onlinecite{Srinivasan_APL08} performed atomistic calculations for Si/SiGe, finding that the presence of different bonds at the interface suppresses $\Delta$. Refs. \onlinecite{Kharche_APL07, Friesen_APL06, Boykin_PRB08} also included models of interface roughness. Nevertheless, EMA studies to date have not considered systematically the effect of interface roughness on the valley-orbit coupling and on the energy-level spectrum of multivalley semiconductor QDs, highlighting the importance of devising a systematic way to quantify interface roughness and study its interplay with the valley degree of freedom.

Interface roughness embodies variations in the position of the interface and variations in the magnitude of the potential step at the interface. Variations in the position of the interface are characterized by two length scales. The rms fluctuations are a measure of the vertical extent of the roughness, while the correlation length contains information about its horizontal spread. The study of roughness began with the theoretical work of Prange and Nee, \cite{Prange_PR68} followed by Ando, \cite{Ando_JPSJ77, Ando_RMP82} yet theoretical and numerical studies have focused mostly on roughness effects on transport. \cite{Stern_APL92, Yamakawa_JAP96, Gamiz_JAP99, Gamiz_JAP03, Mou_PRB00, Croitoru_JAP03} Experimentally Goodnick \textit{et al} \cite{Goodnick_PRB85} used TEM to study roughness at the Si/SiO$_2$ interface, determining the magnitude of the fluctuations and their correlation length. Ourmazd \textit{et al}, \cite{Ourmazd_PRL87} also using TEM, determined that the transition layer between Si and SiO$_2$ spans a distance of approximately 5\AA, in other words approximately two monolayers. The rms fluctuations in the position of the interface are generally found to be of the order of 2-4 \AA,\cite{Ando_RMP82, Goodnick_PRB85} while their correlations extend over a few lattice constants. Experimental studies indicate that state-of-the-art technology can be used to reduce the rms roughness in MOSFETs to the range 0.7-1 \AA.\cite{Sapjeta_MRS07, Yu_MRS05} Later experiments have hinted that the correlation length characterizing roughness may be larger. Yoshinobu \textit{et al} \cite{Yoshinobu_JJAP94} used atomic force microscopy to infer a correlation length of 15 nm, a conclusion upheld by Gotoh \textit{et al} \cite{Gotoh_APL02} using scanning tunneling microscopy, who found a correlation length of 23-26 nm. Even small amplitude rms fluctuations can result in short time scales for mixing valley-split levels, since the wave vectors of the bottoms of the two valleys correspond to very short length scales ($\approx 1\AA$ in Si). All the above works have concentrated on variations in the position of the interface. Variations in the size of the potential step, that is, the \textit{local roughness}, have not been studied to date.

In this article we extend the EMA to study the effect of interface roughness on multivalley QDs.  The approach adopted in this work is equivalent to that of Ref.\ \onlinecite{Saraiva_PRB09}, except that an analytical wave function is used here rather than one determined numerically. Our theory is applicable provided that fluctuations in the location of the interface are much smaller than the spatial extent of the 2DEG, and the fluctuations in the magnitude of the interface potential step are small compared to its average value. We demonstrate that interface roughness has two main effects: it suppresses the valley-orbit coupling and gives rise to tunneling between valley eigenstates in certain dynamical processes in multidot systems. Our starting point is the effective mass approximation in the presence of valleys,\cite{Saraiva_PRB09} with a modified Hamiltonian incorporating interface roughness. Our model of roughness accounts for variations in the position of the interface and variations in the magnitude of the interface potential step. For one electron on one dot (\textit{1e1d}) we study the way interface roughness enters the expression for $\Delta$. If roughness correlations occur on a spatial scale smaller than that of the dot, variations in the position of the interface reduce the magnitude of $\Delta$. For one electron in a double quantum dot (\textit{1e2d}), under these circumstances $\Delta$ is expected to be the same on each dot. Local roughness can cause $\Delta$ to be different on the two dots and induce transitions between valley eigenstates. Interdot transitions between valley eigenstates can be understood as coherent rotations of valley states, and offer the possibility of manipulating valley eigenstates by accessing the appropriate region of the energy spectrum. We subsequently study transitions between two-electron states in a double quantum dot (\textit{2e2d}), which represent the most promising scenario for observing such transitions. For roughness correlations of the order of the size of the DQD, $\Delta$ is not suppressed, yet its phase may differ between adjacent dots. Tunneling between valley eigenstates is enabled in this case. The size of this tunneling matrix element cannot be controlled, although it can be measured experimentally. The findings of this work are general, yet when concrete examples are called for we will discuss QDs made in a Si/SiO$_2$ MOSFET structure, which are actively studied in quantum information. 

It should be pointed out that the long spin coherence time in Si and the possibility of further enhancing the spin coherence time using isotopic purification, which eliminates the $^{29}$Si nuclei, give Si an enormous advantage for quantum information processing.  In addition, the dominant Si-based semiconductor technology allows, in principle, for scalability with the real hope that once a few Si spin qubits are demonstrated in the laboratory, many qubits could be developed without great difficulties. These (i.e. long spin coherence time and scalability) are the reasons that Si spin qubits remain a most active area of research in the context of quantum computation architecture studies.  However, the conduction band valleys in Si, which are degenerate in the simplest effective mass approximation, cause considerable difficulties in fabricating Si spin qubits since the electron spin by itself does not define the state of a Si electron. This is the context in which our detailed theoretical study in this work has been undertaken. The ultimate goal is to help understand the role of valleys in Si spin quantum computation, in particular, to see if the interface roughness naturally present at the Si surface could somehow be utilized to circumvent or even control the Si spin qubit valley degeneracy problem.

The outline of this article is as follows. In Sec.\ \ref{sec:1e1d} we consider the case of a single quantum dot, valleys and valley-orbit coupling. We provide an analytical scheme for calculating the valley-orbit coupling. In Sec.\ \ref{sec:roughness} we introduce a simple model of interface roughness and study the additional terms it introduces into the \textit{1e1d} problem. In Sec.\ \ref{sec:DQD} we introduce a model of a double quantum dot. In Sec. \ref{sec:1e2d} we study intervalley transitions induced by interface roughness for \textit{1e2d} strong tunneling and \textit{1e2d} strong roughness. In Sec.\ \ref{sec:2e2d} we study transitions among \textit{2e2d} singlets and among \textit{2e2d} triplets, while Sec.\ \ref{sec:long} concentrates on the case when interface roughness correlations are of the order of the size of a quantum dot. An extensive discussion of the results, as well as numerical estimates, are contained in Sec.\ \ref{sec:discussion}. We end with a summary of our findings.

\section{Single Quantum Dot}
\label{sec:1e1d}

The EMA assumes the overall potential experienced by the charge carriers is separable and smooth. Conventionally the EMA is understood to apply to physics parallel to the interface, within the 2DEG, as long as the system under study is a few lattice constants away from the interface. The smoothness requirement can be relaxed and the EMA generalized \cite{BenDaniel_PR66} to account for the presence of an interface by considering a $z$-dependent effective mass, smoothly interpolating between the two effective masses on two sides of the interface plane, as we do in this work. In this section we introduce our theoretical model and study a smooth interface as a benchmark for subsequent comparison.

\subsection{Model Hamiltonian}

We study first one electron on one dot. The full Hamiltonian $H_0(x, y, z)$ includes the kinetic and potential energy contributions. The confinement potential is approximated as quadratic in the $xy$-plane, while in the $\hat{\bm z}$-direction one has the interface potential $\mathcal{V} (z)$ and electric field $F$
\begin{equation}
V_D (x, y, z) = \frac{\hbar^2}{2m^*a^2} \, \bigg[ \frac{(x - x_D)^2 + y^2}{a^2} \bigg] + \mathcal{V} (z) + eFz.
\end{equation}
The dot is located at $(x_D, 0, 0)$ and has a Fock-Darwin radius $a$. The interface potential $\mathcal{V}$ will be discussed in Sec.\ \ref{sec:roughness}. Within the effective mass approximation (EMA) the wave functions for the quantum dot are written as $D_\xi (x, y, z) = \phi_D(x,y) \, \psi (z) \, u_\xi ({\bm r})\, e^{ik_\xi z}$, where, as in our previous work, we have introduced a valley index $\xi = \{ z, \bar{z} \}$ that will become relevant below, and $k_{z, \bar{z}} = \pm k_0 \equiv \pm 0.85 \, (2\pi/a_{Si})$, with $a_{Si}$ the lattice constant of Si.\cite{Culcer_PRB09, Culcer_10} For the envelopes $\phi_D (x, y)$ we use the Fock-Darwin states
\begin{equation}
\phi_D (x, y) = \frac{1}{a\sqrt{\pi}} \, e^{-\frac{(x - x_D)^2 + y^2}{2a^2}},
\end{equation}

The effective mass equation for motion in the $z$-direction, perpendicular to the interface, is
\begin{equation}
\bigg[-\frac{\hbar^2}{2m_z} \, \frac{\partial^2}{\partial z^2} + \mathcal{V} (z) + eFz\bigg] \psi(z) = \varepsilon_z \psi(z).
\end{equation}
The effective mass for SiGe is the bare electron mass $m_0$, while for SiO$_2$ it is found to range between $0.3$ and $0.4m_0$.\cite{Brar_SiO2_TnlMeff_APL96, Depas_SiO2TnlMeff_SSE95} For $\psi(z)$ we use a modified Fang-Howard variational wave function. \cite{Bastard} With the interface at $z=0$
\begin{equation}
\begin{array}{rl}
\displaystyle \psi(z) = & \displaystyle M \, e^{\frac{k_b z}{2}}, z < 0 \\ [3ex]
= & \displaystyle N\, (z + z_0) \, e^{\frac{-k_{Si} z}{2}}, z > 0.
\end{array}
\end{equation}
We will also use $b = 1/k_{Si}$ as a length scale characteristic of the size of the 2DEG along the growth direction. Continuity of the wave function at the interface requires $M = Nz_0$, and that of $(1/m_z)\, d\psi/dz$ gives
\begin{equation}
N = \frac{1}{\sqrt{\frac{2}{k_{Si}^3} + \frac{2z_0}{k_{Si}^2} + z_0^2 \big(\frac{1}{k_{Si}} + \frac{1}{k_b}\big)}}.
\end{equation}
The parameter $z_0$ is given by 
\begin{equation}
z_0 = \frac{2}{k_{Si} + k_b\big(\frac{m_{Si}}{m_b}\big)}
\end{equation}
In this work $k_b$ is fixed at $k_b = \sqrt{\frac{2m_bU_0}{\hbar^2}}$, leaving $k_{Si}$ as the only variational parameter. The energy $\varepsilon_z$ that is to be minimized is given by
\begin{widetext}
\begin{equation}
\varepsilon_z = \frac{\hbar^2N^2}{8m_z b} \, (2b^2 + 2bz_0 - z_0^2) + \frac{N^2 eF}{2} \, b^2 \, (6b^2 + 4bz_0 + z_0^2) - \frac{\hbar^2M^2k_b}{8m_b b} + \frac{M^2 U_0}{k_b} - \frac{M^2eF}{2k_b^2},
\end{equation}
\end{widetext}
which includes a factor of $1/2$ in the electric field term to account for double counting of electrons. \cite{Bastard} 

\subsection{Valley-orbit coupling for a sharp flat interface}
\label{sec:VOC}

The conduction band of some materials consists of a series of equivalent valleys. In such circumstances, for a smooth potential the effective mass equation has two equivalent solutions. If the potential is not smooth the solutions could be mixed, leading to valley-orbit coupling. Consider the basis states $\{ D_\xi \}$, where $D = L, R$ represents either of the left and right dots. The $z$ and $\bar{z}$ states have a vanishingly small overlap, which is neglected here. With this approximation these states are orthogonal. To determine the one-dot energy levels we require first the matrix elements of the Hamiltonian in the basis spanned by $\{ D_\xi \} = $ either $\{ L_\xi \}$ or $\{ R_\xi \}$
\begin{equation}\label{H1e1d}
H_{1e1d} = \varepsilon_D + \begin{pmatrix} 0 & \Delta_D \cr \Delta_D^* & 0 \end{pmatrix},
\end{equation}
where $\varepsilon_D$ is the confinement energy, which is large, thus we consider only the lowest level $\varepsilon_D \equiv \varepsilon^{(0)}_D$, generally of the order of tens of meV. The lowest energy level for each dot consists of two valleys connected by a valley-orbit coupling $\Delta_D = |\Delta_D| \, e^{-i\phi_D}$. The eigenstates of $H_{1e1d}$ can be expressed as
\begin{equation}
\ket{D_\pm} = \frac{1}{\sqrt{2}} \, (\ket{D_z} \pm e^{i\phi_D} \ket{D_{\bar{z}}}).
\end{equation}
In this work \textit{intervalley} refers to matrix elements connecting the $z, \bar{z}$ states, whereas \textit{between valley eigenstates} refers to matrix elements connecting $+$ and $-$ states. The valley eigenstates are the same for both dots if the interface is sharp along the growth direction and flat perpendicular to it. The valley-orbit coupling is represented by the matrix element \cite{Saraiva_PRB09}
\begin{equation}
\Delta_D = \tbkt{D_z}{(\mathcal{V} + eFz)}{D_{\bar{z}}}.
\end{equation}
Its magnitude gives the valley splitting $|\Delta_D|$. The dominant contribution to $\Delta_D$ comes from the interface potential, while the contribution due to $F$ has been found to be negligible.\cite{Saraiva_PRB09} For a perfectly smooth and perfectly sharp interface $\Delta_D$ is equal to
\begin{widetext} 
\begin{equation}\label{Delta}
\begin{array}{rl}
\displaystyle \Delta_D = & \displaystyle U_0 \, \int \int dx\, dy\, |\phi(x,y)|^2  \int_{-\infty}^{\infty} dz\, \Theta(-z)\, |\psi (z)|^2 \, e^{-2ik_z z}\, u^*_z ({\bm r}) \, u_{\bar{z}} ({\bm r}) = U_0 \, N^2 \, z_0^2 \sum_{{\bm K}, Q_z} \, \frac{c^{z*}_{{\bm K}} c^{\bar{z}}_{{\bm K} + Q_z{\bm z}}}{q_z},
\end{array}
\end{equation}
where $q_\xi = k_b + iQ_z - 2ik_\xi$. The matrix element between the same valley states $\Lambda_D = \tbkt{D_\xi}{\mathcal{V}}{D_\xi}$ is given by 
\begin{equation}\label{Lambda}
\Lambda_D = U_0 \, \int \int dx\, dy\, |\phi(x,y)|^2  \int dz\, \Theta(-z)\, |\psi (z)|^2 \,  |u_\xi ({\bm r})|^2 = U_0 \, N^2 \, z_0^2 \sum_{{\bm K}, Q_z} \, \frac{c^{\xi*}_{{\bm K}} c^{\xi}_{{\bm K} + Q_z{\bm z}} }{\beta_\xi},
\end{equation}
\end{widetext}
where $\beta_\xi = k_b + iQ_z$. We have used the general notation $\xi$ for the valley indices in $\Lambda_D$ since we anticipate $\tbkt{D_z}{\mathcal{V}}{D_z} = \tbkt{D_{\bar{z}}}{\mathcal{V}}{D_{\bar{z}}} $. Detailed derivations of Eqs. (\ref{Delta}) and (\ref{Lambda}) are presented in Appendix \ref{app:VOC}. The coefficients $c^\xi_{{\bm K}}$ for Si were determined in Ref.\ \onlinecite{Koiller_PRB04}. Using these values, we find that by far the biggest contributions to $\Delta_D$ and $\Lambda_D$ come from the terms with $Q_z = 0$. Given that the magnitude of $|k_b - 2ik_0| \approx 2k_0$, it follows that $\Delta_D$ can be approximated simply by $\Delta_D \approx iU_0 \, \Sigma \, N^2 \, z_0^2/(2k_0)$, where $\Sigma = \sum_{\bm K} \, c^{\xi*}_{\bm K} c^{-\xi}_{\bm K}$. The Umklapp terms, which have $Q_z = 2\xi \, (2\pi/a_{Si} )$, give imaginary contributions and sum to zero. We will refer to the valley-orbit coupling arising from the sharp flat interface as the \textit{global} valley-orbit coupling $\Delta_0$, without the dot subscript, and similarly for the global $\Lambda_D = \Lambda_0$. 

To obtain numerical estimates of $|\Delta_0|$ in Si/SiO$_2$ and Si/SiGe an interfacial electric field of 150 kV/cm is assumed, which is the same value that was used in Ref. \ \onlinecite{Saraiva_PRB09}, in order to enable a direct comparison. For a Si/SiO$_2$ interface, with $U_0 \approx 3$eV (as discussed in Ref. \ \onlinecite{Ando_RMP82}) and $m_b = 0.4 m_0$, $b$ is optimized at 1.058 nm, for which we find $|\Delta_0| \approx$0.11 meV. Since the effective mass is not well defined for SiO$_2$, we have also used $m_b = 0.3 m_0$, finding $b \approx 1.01$nm and $|\Delta_0| \approx$0.08 meV. For a Si/SiGe interface, with $U_0 \approx 150$meV, $b$ is optimized at 1.215 nm, whence $|\Delta_0| \approx$0.1 meV. Considering the relatively simple variational function that is used in this work, the comparison with accurate numerics \cite{Saraiva_PRB09} is extremely encouraging. The real and imaginary parts of $\Delta_0$ are given to a very good approximation by $k_b/\sqrt{k_b^2 + 4k_0^2}$ and $2k_0/\sqrt{k_b^2 + 4k_0^2}$. Consequently, with $k_0$ fixed, as the interface potential step becomes stronger the imaginary part of $\Delta_0$ also increases. For the Si/SiO$_2$ interface the ratio of the imaginary and real parts of $\Delta_0$ is found to be 3.5, while for the Si/SiGe interface it is 15.7. \footnote{For the real and imaginary parts of the valley-orbit coupling a direct comparison with Ref.\ \onlinecite{Saraiva_PRB09} cannot be made, since in that work the phase of the Bloch functions needs to be specified, leading to ambiguity in the overall phase of $\Delta_0$.} The ratio of the off-diagonal matrix element $\Delta_0$ of the interface potential to its diagonal matrix element $\Lambda_0$ is given approximately by $k_b/(k_b - 2ik_0)$, and the magnitude of this ratio is 0.28 for the Si/SiO$_2$ interface and 0.06 for the Si/SiGe interface. The final result varies significantly as a function of $b$. \footnote{We have used modified Fang-Howard wave function due to the physical insight it offers, and relatively transparent analytical results. Slightly better quantitative results may be found analytically using the wave function of Ref.\ \onlinecite{Sham_PRB79}, yet the results cannot be expressed in closed form and the generalization to allow penetration into the insulator is not obvious.}

\section{Interface roughness}
\label{sec:roughness}

An interface may be observed to be sharp and of a high quality, yet the transition layer between two materials still always has a finite width. The atoms on the two sides of the interface have different bonding, as do the atoms at the interface. For example between Si on one side and SiO$_2$ on the other there exists a region a few monolayers thick composed of SiO$_x$, with $0 \le x \le 2$.\cite{Ourmazd_PRL87} Consequently, even when the interface is sharp the finite transition layer alters the energy landscape. \footnote{See Fig. 43 of Ref. \onlinecite{Ando_RMP82} for a model of the interface.} Experimentally roughness can be detected by transmission electron microscopy,\cite{Goodnick_PRB85, Ourmazd_PRL87} atomic force microscopy, \cite{Yoshinobu_JJAP94} scanning tunneling microscopy,\cite{Gotoh_APL02} and other surface characterization techniques such as Auger spectroscopy, angle-resolved photoemission spectroscopy, low-energy electron diffraction and resonant high-energy electron diffraction.\cite{Luth}

The notion of a sharp barrier breaks down at the atomic scale. The ideal method for dealing with this issue is to use the exact microscopic structure of the interface. Such a treatment is only possible numerically, yet even then the interface structure is not known exactly and, when one of the materials is amorphous, it can only be dealt with approximately and/or statistically. Therefore, there exists a scope for gaining analytical insight by refining the effective mass approximation to include the effects of interface roughness. Within the EMA the interface is modeled in such a way that the atomic structure is smoothed out, and an effective mass appropriate for the insulator is used. In the same way as the smoothness requirement can be relaxed as long as the region under study is a few monolayers away from the interface, one can also relax the requirement that the interface be sharp and flat. Corrections to this picture can be made so that the location of the interface varies on an atomic scale but with an amplitude much smaller than the effective size of the confinement. Then the $xy$-motion and the $z$-motion are still approximately separable. Beyond this refinement, a random potential is added to allow the \textit{magnitude} of the interface step to fluctuate. These two features can be regarded as corrections to the effective mass picture, which is entirely based on bulk parameters.

A simple model, which was introduced by Prange and Nee \cite{Prange_PR68} and developed by Ando,\cite{Ando_JPSJ77, Ando_RMP82} has become the standard model used in studies of surface roughness. Within this model the interface is described by a potential $\mathcal{V} [z - \zeta(x,y)]$, in which the fluctuating position of the interface $\zeta(x,y)$ is treated as a random function of $x$ and $y$. Prange and Nee\cite{Prange_PR68} assumed that the probability for an electron to be specularly reflected at the interface is nearly unity, with a boundary condition that the wave function vanish at the interface. Ando\cite{Ando_JPSJ77, Ando_RMP82} approached the problem by expanding $\mathcal{V}$ in $\zeta$ and retaining the term linear in $\zeta$. This approach may be satisfactory for calculating the scattering matrix element due to roughness. Yet, as Refs.\ \onlinecite{Ando_JPSJ77, Ando_RMP82} themselves clearly state, there is no rigorous justification for this expansion. As we will show in this work, indeed this expansion is not at all justified when the effect of roughness on the valley-orbit coupling is required. The breakdown of the perturbative expansion is most emphatically manifest in Section \ref{sec:long}, when long interface correlations are considered, yet the expansion is also not valid if the $\zeta$-correlations span a small spatial scale, as Section \ref{sec:short} demonstrates.

\subsection{Model of roughness}
\label{sec:model}

We treat interface roughness using a slightly different approach from that of Refs. \onlinecite{Ando_JPSJ77} and \onlinecite{Ando_RMP82}. We consider an interface potential given by 
\begin{equation}
\begin{array}{rl}
\displaystyle \mathcal{V} (x,y,z) = & \displaystyle [U_0 + V(x,y)] \, f[z - \zeta(x,y)].
\end{array}
\end{equation}
where $U_0 \gg V$ for all $x$ and $y$. We take a general function $f(z - \zeta)$ to describe the interface, where $f(z)$ is a well-defined function of $z$, while $\zeta$ is a random function of $x$ and $y$ accounting for the variation in the location of the interface.  For an interface that is perfectly sharp in the $\hat{\bm z}$-direction and perfectly smooth in the $xy$-plane (the ideal interface), one can write $\mathcal{V} (z) = U_0 \, \Theta(-z)$. For an interface whose location fluctuates in the $xy$ plane, but is still sharp, one can use $f(z - \zeta) = \Theta[-(z - \zeta)]$. The potential $V(x,y)$ represents the local roughness, which gives fluctuations in the magnitude of the potential step at the interface. We treat $V(x,y)$ perturbatively, since we expect $V(x,y) \ll U_0$, but we do not do perturbation theory in $\zeta(x, y)$. In this problem the criterion for smallness is generally nontrivial. The discussion of this criterion will follow naturally after the analytical expressions for the effect of the roughness are presented.

Aside from variations in the location of the interface, sources of roughness include dangling bonds, impurities, and variations in bond length in the vicinity of the interface. All of these contribute to the local roughness $V(x,y)$. We expect, however, that variations in bond length will occur on the atomic length scale, which is orders of magnitude smaller than the spatial extent of a quantum dot. This implies that the effect of bond length fluctuations averages to zero over the quantum dot and is washed out. We do not consider it explicitly in what follows. 

\subsection{Applicability of the EMA}

The location of the interface $\zeta$ is a \textit{random} function of $x$ and $y$. It has an average value $\bkt{\zeta} = 0$, while its variance will be denoted simply by $\bkt{\zeta^2}$. The notation $\bkt{}$ denotes the average over the $xy$-plane, while $\bkt{}_{QD}$ denotes the average over $x$ and $y$ spanning the region occupied by the quantum dot $\equiv 1/(\pi a^2) \int\int dx\, dy\, e^{-\frac{x^2 + y^2}{a^2}}$. It is assumed that the fluctuations around $\bkt{\zeta}$ are small compared to $b$, which represents the spatial scale over which the wave function changes by a significant amount in the $\hat{\bm z}$-direction (i.e. the size of the confinement.) The spatial scale of the roughness divided by the spatial scale over which the wave function changes significantly is an important parameter. The size of this parameter in comparison to unity determines the applicability of the EMA. It is necessary for $\zeta (x,y)$ to be small, so that the potential acting on the charge carriers remains approximately separable and use of the effective mass approximation is justified. If $|\zeta| \ll a,b $ the fluctuation in the $z$-direction is considerably less than the spatial extent in all three directions, so that the wave function is still approximately separable in $x, y, z$. This condition must be augmented by a formal criterion according to which the potential is weak. In principle all applied electric fields, such as $F$ in this problem, violate the criterion for being weak, since the potential $eFz$ eventually becomes large on any scale relevant to the problem once $z$ exceeds a certain threshold. This fact however is not usually a problem as long as the potential $\mathcal{V}$ does not vary too fast on lattice length scales, that is, as long as $eF a_{Si}$ is significantly smaller than the average atomic potential in a unit cell. To summarize, the EMA should remain valid as long as the random potential $V(x,y) \ll U_0$ for all $x,y$ and the fluctuations in the location of the interface are much smaller than the spatial extent of the 2DEG.

\subsection{Roughness effect on the valley-orbit coupling}

We use the notation $\Delta_0$ for the unperturbed valley-orbit coupling in the absence of interface roughness, which is the same on both dots. The valley-orbit coupling in the presence of roughness is denoted by $\Delta$. We require $ |\Delta - \Delta_0| \ll |\Delta_0|$ for any $x,y$. In the presence of an interface located at $z = \zeta$ the wave function must be altered as $\psi(z) \rightarrow \psi(z - \zeta)$. The matrix elements of the interface potential are found in the same way as before
\begin{widetext} 
\begin{equation}
\begin{array}{rl}
\displaystyle \Delta_D = & \displaystyle \int\int dx\, dy\, (U_0 + V) \,  |\phi(x,y)|^2  \int dz\, \Theta[-(z - \zeta)] \, |\psi (z - \zeta)|^2 \, e^{-2ik_z z}\, u^*_z ({\bm r}) \, u_{-z} ({\bm r}) = \frac{N^2 \, z_0^2 \, \bkt{(U_0 + V)  \, e^{i \kappa_z \, \zeta}}_{QD}}{q_z},
\end{array}
\end{equation}
\end{widetext}
The effect of the interface is contained in two terms: the spatial average of the exponential $e^{i \kappa_\xi \, \zeta}$, and the spatial average of the product $V(x, y) \, e^{i \kappa_\xi \, \zeta}$. We recall that in Si the contribution to $\Delta$ due to one wave vector, namely $Q_z = 0$, overwhelms all the others. Therefore a good indication of the effect of interface roughness is given by the averages $\bkt{e^{2 i \,k_0\, \zeta}}_{QD}$ and $\bkt{V(x,y) \, e^{2 i \,k_0\, \zeta}}_{QD}$. The appearance of the simple exponential oscillatory terms such as $e^{i \kappa_\xi \, \zeta}$ is a consequence of the fact that we are using the effective mass approximation. In a more complicated numerical treatment we still expect a strong dependence on $\kappa_\xi \, \zeta$, though it will likely not be expressible as a simple exponential. Since it is predominantly the term with $Q_z = 0$ that contributes to $\Lambda_D$, this quantity should not be affected by the presence of $\zeta$. This can be seen immediately by considering 
\begin{widetext} 
\begin{equation}
\begin{array}{rl}
\displaystyle \Lambda_D = & \displaystyle \int\int dx\, dy\, (U_0 + V) \,  |\phi(x,y)|^2  \int dz\, \Theta[-(z - \zeta)] \, |\psi (z - \zeta)|^2 \, |u_\xi ({\bm r})|^2 = \frac{N^2 \, z_0^2 \, \bkt{(U_0 + V)}_{QD}}{\beta_\xi}.
\end{array}
\end{equation}
\end{widetext}

We distinguish between $V(x,y)$, which can be treated perturbatively, and $e^{i \kappa_\xi \, \zeta}$, which is non-perturbative.  The most suitable approach to the problem is determined by the spatial scales involved. Experiments indicate that the spatial scales characterizing $\zeta$ are $2-4\AA$ for the rms and $5-15\AA$ for the correlation length. Several studies have indicated that the roughness in Si/SiO$_2$ could be reduced, to an rms of the order of $\approx$ 1 \AA \cite{Yu_MRS05} and even as far as 0.7\AA.\cite{Sapjeta_MRS07} At the same time, a series of experiments using scanning tunneling microscopy suggested that the horizontal correlations of $\zeta$ could stretch over distances of the order of 30nm. We will use \textit{short} correlations to refer to spatial correlations of $\zeta$ on a scale much smaller than the typical extent of a quantum dot (approximately 50nm), and \textit{long} correlations for spatial correlations of $\zeta$ on a scale of the order of or greater than the typical extent of a quantum dot. The problem of valley-orbit coupling is vastly different for short interface correlations and long interface correlations. Physically, fluctuations associated with a random interface ought to occur on the scale of the lattice constant. We anticipate that the true physical picture is represented by short correlations, but given the existing uncertainty we will study both cases in this paper.

In the most general case the valley-orbit coupling may be different on the two dots. Since the valley-orbit coupling is a complex number, in principle one should expect both the amplitude and the phase to be different between the two dots. This fact is not of much significance when a sharp flat interface is considered, yet is of enormous importance when interface roughness is concerned. In addition, roughness is expected to show significant sample dependence. In double quantum dots in different samples the roughness profile on the left dot may be qualitatively similar to or very different from the roughness profile on the right dot. One also expects a significant difference between SiO$_2$ and SiGe, with the former generally having larger interface roughness. From this perspective it will be useful to be able to compare e.g. gate-defined quantum dots with disordered quantum dots.

In addition, we have noted that, in sums of the type $\sum_{\bm G} c_{\bm G} e^{i{\bm G}\cdot{\bm r}}$, the only sizable terms correspond to the reciprocal lattice vector ${\bm G} = (000)$. We infer that terms with wave vectors larger than that, corresponding to spatial scales of $1\AA$ or less (\textit{atomic scale roughness}), should not be important in determining $\Delta_D$ and $\Lambda_D$. This reinforces the observation of Sec. \ref{sec:model} concerning variations in bond length.

\section{Double quantum dot}
\label{sec:DQD}

In a single quantum dot interface roughness affects the valley-orbit coupling. In a double quantum dot it affects in addition the relative values of the dot-averaged valley-orbit splitting $\Delta_D$ on the two dots, as well as interdot tunneling, potentially giving rise to interesting electron dynamics. The confinement of a double quantum dot is modeled by a biquadratic potential
\begin{equation}
\arraycolsep 0.3 ex
\begin{array}{rl}
\displaystyle V_{DQD} (x, y, z) & \displaystyle  = \frac{\hbar \omega_0}{2} \, \{ \mathrm{Min} [\frac{(x-X_0)^2}{a^2}, \frac{(x+X_0)^2}{a^2}] + \frac{y^2}{a^2} \} \\ [3ex]
& \displaystyle - eEx + eFz + \mathcal{V}(z).
\end{array}
\end{equation}
As in Ref. \ \onlinecite{Culcer_10}, we orthogonalize the wave functions $\{ D_\xi \}$ to get the Wannier functions $\{ \tilde{D}_\xi \}$, given by $\tilde{L}_\xi = \frac{L_\xi - gR_\xi}{\sqrt{1 - 2lg + g^2}}$ and $\tilde{R}_\xi = \frac{R_\xi - gL_\xi}{\sqrt{1 - 2lg + g^2}}$, where $g = (1 - \sqrt{1 - l^2})/l$ and $l =\langle L_\xi | R_\xi \rangle$. This ensures that $\langle \tilde{R}_\xi | \tilde{L}_\xi \rangle = 0$. One diagonalizes the Hamiltonians $\tilde{H}_D$ and determines the eigenfunctions $\{ \tilde{D}_\pm \}$. We will consider one electron initialized in a DQD (\textit{1e2d}) as in a QD charge qubit, as well as two electrons (\textit{2e2d}), which can form singlet and triplet states (given in Ref. \onlinecite{Culcer_10}.) Henceforth we work in this basis, and to indicate this fact we augment all quantities with a tilde, thus $\tilde{\Delta}$ etc.

Our principal assumption in setting up the problem is that the physics specific to the valleys can be considered separately for the left and right dots. The starting point of our approach envisages two dots and one electron primarily localized on one of the dots. The valley degree of freedom implies that there are two possible single-particle states for each dot, while the interface potential gives a coupling between the two valleys on each dot. We diagonalize the single-particle Hamiltonian separately for the left and right dots, and obtain the valley splitting and valley eigenstates for each dot. We neglect matrix elements of the interface potential connecting the left and right dots, which represent terms of order $l\Delta$. Under these circumstances $\phi_D$ is not a function of position, a fact that provides a crucial simplifying assumption in our algebra. 

The assumption can be recast into the statement that there is no interdot tunneling between different valleys, meaning that the tunneling matrix element only connects states from the same valley. One can tunnel between $\ket{L_\xi}$ and $\ket{R}_\xi$, but not $\ket{L_\xi}$ and $\ket{R}_{-\xi}$. Tunneling between valley eigenstates is determined only by the phases of each valley wave function in the valley eigenstates on each dot, which in turn are determined only by the details of the interface potential in the region of the dot.

A full treatment of the \textit{1e 2d} and \textit{2e2d} problems would take as its starting point the four valley wave functions $\{\ket{L_\xi}\}$ and $\{\ket{R}_\xi\}$. It takes into account all the matrix elements of the interface potential, including matrix elements of the form $\tbkt{L_\xi}{\mathcal{V}}{R_{-\xi}}$. In this case the eigenstates are mixtures of all four functions $\{\ket{L_\xi}\}$ and $\{\ket{R}_\xi\}$, a far longer and more cumbersome approach. Since for a typical DQD $l \ll 1$ by several orders of magnitude, terms of order $l\Delta$ are unlikely to play an important role, and it is reasonable to anticipate that our simplified approach contains the dominant physics. Given that $\tilde{\Delta}$ differs from $\Delta$ by terms of order $l\Delta$, which are neglected here, we can consider $\tilde{\Delta} \approx \Delta$.


\subsection{One-electron Hamiltonian}

The most general Hamiltonian for one electron in a DQD in the basis $\{ \tilde{L}_+, \tilde{L}_-, \tilde{R}_+, \tilde{R}_-\}$ is
\begin{equation}
H_{1e2d} = \tilde{\varepsilon}^L + \begin{pmatrix}
|\tilde{\Delta}^L_{+}| & v^L_{+-} & \tilde{t}_{0++} & \tilde{t}_{0+-}  \cr
v^L_{+-} & - |\tilde{\Delta}^L_{-}| & \tilde{t}_{0-+} & \tilde{t}_{0--} \cr
\tilde{t}_{0++} & \tilde{t}_{0+-} & - \delta + |\tilde{\Delta}^R_{+}| & v^R_{+-} \cr
\tilde{t}_{0-+} & \tilde{t}_{0--} & v^R_{-+} & - \delta - |\tilde{\Delta}^R_{-}| 
\end{pmatrix},
\end{equation}
where the detuning is defined as $\delta = \tilde{\varepsilon}^L - \tilde{\varepsilon}^R$, and $|\tilde{\Delta}^D_\pm| = |\tilde{\Delta}| \pm \tilde{v}^D_{\pm\pm}$. The tunneling parameter $\tilde{t}_{0+-}$ is defined as $\tilde{t}_{0+-} = \tbkt{\tilde{L}_+}{H}{\tilde{R}_-}$. We will use the notation $\tilde{v}^D_{\xi\xi'} = \tbkt{\tilde{D}_\xi}{V(x,y)}{\tilde{D}_{\xi'}}$, with corresponding notation for the $\pm$ states. Throughout the remainder of this article, we will also use $\tilde{v}$ to refer (generally) to the magnitude of the off-diagonal term $|\tilde{v}^D_{+-}|$, an important parameter in determining several regimes of qualitatively different physics.

\subsection{Two-electron Hamiltonian}
\label{sec:2eHam}

For \textit{2e2d}, the interdot tunneling parameter $\tilde{t}$ has a single-particle part $\tilde{t}_0$ and a Coulomb-interaction (enhancement) part $\tilde{s}$,\cite{Burkard_PRB99, Qiuzi_PRB10, Culcer_PRB09} the total being $\tilde{t} = \tilde{t}_0 + \tilde{s}$, with $\tilde{s} = \tbkt{L_\pm^{(1)} L_\pm^{(2)}}{V_{ee}}{L_\pm^{(1)}  R_\pm^{(2)}} = \tbkt{L_\pm^{(1)}  L_{\mp}^{(2)}}{V_{ee}}{L_\pm^{(1)}  R_{\mp}^{(2)}}$.\cite{Culcer_10} The Coulomb exchange term $\tilde{j}$ is much smaller than the other terms appearing in the Hamiltonian and is not needed at all in this paper. We also define the critical detuning as $\delta_c = (\tilde{u} - \tilde{k})$, where $\tilde{u}$ represents the on-site Coulomb interaction and $\tilde{k}$ the direct Coulomb interaction between electrons on the two dots. \cite{Culcer_PRB09, Culcer_10} The effective detuning is defined as $\tilde{\delta} = \delta - \delta_c$, measured from this critical detuning. In the basis $\{\tilde{S}^{LR}_{- -}, \tilde{S}^{RR}_{- -}, \tilde{S}^{LR}_{++}, \tilde{S}^{RR}_{+ +}, \tilde{S}^{LR}_{+-}, \tilde{S}^{LR}_{-+}, \tilde{S}^{RR}_{+-}\}$ used in Ref. \onlinecite{Culcer_10} the most general singlet Hamiltonian is $\tilde{H}_S = \tilde{H}^{(1)}_S + \tilde{V}_S$, where
\begin{widetext}
\begin{equation}
\arraycolsep 0.3 ex
\begin{array}{rl}
\displaystyle \tilde{H}^{(1)}_S = \displaystyle \tilde{\varepsilon}_L + \tilde{\varepsilon}_R + \tilde{k} + \begin{pmatrix}
- 2|\tilde{\Delta}| & \tilde{t}_{++} \sqrt{2} & 0 & 0 & 0 & 0 & \tilde{t}_{+-} \cr\cr
\tilde{t}_{++}\sqrt{2}  & - \tilde{\delta} - 2 |\tilde{\Delta}| & 0 & 0 & 0 & \tilde{t}^*_{-+}\sqrt{2} & 0 \cr\cr
0 & 0 & 2|\tilde{\Delta}| &  \tilde{t}_{++} \sqrt{2} & 0 & 0 & \tilde{t}_{-+} \cr\cr
0 & 0 &  \tilde{t}_{++} \sqrt{2} & - \tilde{\delta} + 2 |\tilde{\Delta}| & \tilde{t}^*_{+-}\sqrt{2} & 0 & 0 \cr\cr
0 & 0 & 0 & \tilde{t}_{+-}\sqrt{2} & 0 & 0 &  \tilde{t}_{++} \cr\cr
0 & \tilde{t}_{-+}\sqrt{2} & 0 & 0 &  0 & 0 & \tilde{t}_{++}  \cr\cr
\tilde{t}_{+-}^* & 0 & \tilde{t}_{-+}^* & 0 & \tilde{t}_{++} & \tilde{t}_{++} & - \tilde{\delta}
\end{pmatrix}
\end{array}
\end{equation}
$\tilde{H}^{(1)}_S$ was given in Ref.\ \onlinecite{Culcer_10}without the tunneling matrix element $\tilde{t}_{+-} $. $\tilde{V}_S$ represents the matrix elements of the local roughness between singlet wave functions, which are discussed in Appendix \ref{app:matrixel}. It is given by
\begin{equation}
\arraycolsep 0.3 ex
\begin{array}{rl}
\displaystyle \tilde{V}_S =  \begin{pmatrix}
\tilde{v}^{L}_{--} + \tilde{v}^{R}_{--} & 0 & 0 & 0 & \tilde{v}^L_{-+} & \tilde{v}^R_{-+} & 0 \cr\cr
0  & 2\tilde{v}^{R}_{--} & 0 & 0 & 0 & 0 & \sqrt{2} \, \tilde{v}^R_{-+} \cr\cr
0 & 0 & \tilde{v}^{L}_{++} + \tilde{v}^{R}_{++} & 0 & \tilde{v}^R_{+-} & \tilde{v}^L_{+-} & 0 \cr\cr
0 & 0 & 0 & 2\tilde{v}^{R}_{++} & 0 & 0 & \sqrt{2} \, \tilde{v}^R_{+-}  \cr\cr
\tilde{v}^L_{+-} & 0 & \tilde{v}^R_{-+} & 0 & \tilde{v}^{L}_{++} + \tilde{v}^{R}_{--} & 0 & 0 \cr\cr
\tilde{v}^R_{+-} & 0 & \tilde{v}^L_{-+} & 0 &  0 & \tilde{v}^{L}_{--} + \tilde{v}^{R}_{++} & 0  \cr\cr
0 & \sqrt{2} \, \tilde{v}^R_{+-}  & 0 & \sqrt{2} \, \tilde{v}^R_{-+} & 0 & 0 & \tilde{v}^{R}_{++} + \tilde{v}^{R}_{--} 
\end{pmatrix}
\end{array}
\end{equation}
The most general triplet Hamiltonian in the basis $\{\tilde{T}^{LR}_{- -}, \tilde{T}^{LR}_{++}, \tilde{T}^{LR}_{+-}, \tilde{T}^{LR}_{-+}, \tilde{T}^{RR}_{+-}\}$ used in Ref. \onlinecite{Culcer_10} is $\tilde{H}_T = \tilde{H}_T^{(1)} + \tilde{V}_T$, where
\begin{equation}
\arraycolsep 0.3 ex
\begin{array}{rl}
\displaystyle \tilde{H}_T ^{(1)}= & \displaystyle \tilde{\varepsilon}_L + \tilde{\varepsilon}_R + \tilde{k} + \begin{pmatrix}
- 2 |\tilde{\Delta}| & 0 & 0 & 0 & -\tilde{t}_{+-} \cr
0 & 2|\tilde{\Delta}| & 0 & 0 & \tilde{t}_{-+} \cr
0 & 0 & 0 & 0 & \tilde{t}_{++} \cr
0 & 0 &  0 & 0 & - \tilde{t}_{++} \cr
-\tilde{t}_{+-}^* & \tilde{t}_{-+}^* & \tilde{t}_{++} & - \tilde{t}_{++} & - \tilde{\delta}
\end{pmatrix}
\end{array}
\end{equation}
and 
\begin{equation}
\arraycolsep 0.3 ex
\begin{array}{rl}
\displaystyle \tilde{V}_T =  \begin{pmatrix}
\tilde{v}^{L}_{--} + \tilde{v}^{R}_{--} & 0 & \tilde{v}^L_{-+} & \tilde{v}^R_{-+} & 0 \cr\cr
0 & \tilde{v}^{L}_{++} + \tilde{v}^{R}_{++} & \tilde{v}^R_{+-} & \tilde{v}^L_{+-} & 0 \cr\cr
\tilde{v}^L_{+-} & \tilde{v}^R_{-+} & \tilde{v}^{L}_{++} + \tilde{v}^{R}_{--} & 0 & 0 \cr\cr
\tilde{v}^R_{+-} & \tilde{v}^L_{-+} &  0 & \tilde{v}^{L}_{--} + \tilde{v}^{R}_{++} & 0  \cr\cr
0 & 0 & 0 & 0 & \tilde{v}^{R}_{++} + \tilde{v}^{R}_{--}
\end{pmatrix}
\end{array}
\end{equation}
\end{widetext}
The matrix elements of $V_T$ are discussed in Appendix \ref{app:matrixel}.

If one assumes that the valley-orbit coupling has the same form in both quantum dots then both have the same valley splitting and valley eigenstates. When electrons tunnel between the dots they tunnel between like valley eigenstates, and interdot intervalley tunneling is strongly suppressed. In other words $\tilde{t}_{+-} = 0$ above. In the absence of any interface roughness, with $\tilde{t}_{+-} = \tilde{v} = 0$, the spectrum of two electrons in a DQD can be clearly divided into three branches with different valley composition,\cite{Culcer_10} which may be labeled by $++, - -, +-$. These branches may be mixed by interface roughness. As we will discuss in the sections below, two features of this mixing are noticeable. Firstly, roughness only mixes certain branches of the spectrum. There is no mixing between the $++$ and $--$ branches, but both $++$ and $--$ are mixed by roughness with $+-$. Moreover, there is no mixing of states within any of the $++/--/+-$ subspaces by the local roughness. Secondly, in situations in which $\tilde{t}_{+-}$ is nonzero (see Sec.\ \ref{sec:long}), $\tilde{t}_{+-}$ and $\tilde{v}$ couple different states: no states are coupled by both $\tilde{v}$ and $\tilde{t}_{+-}$ for one or two electrons in a DQD. Therefore, quite generally, variations in the magnitude of the potential step and variations in the location of the interface have qualitatively different effects on interdot electron dynamics.

\section{Short interface correlations}
\label{sec:short}

\begin{figure}[tbp]
\includegraphics[width=\columnwidth]{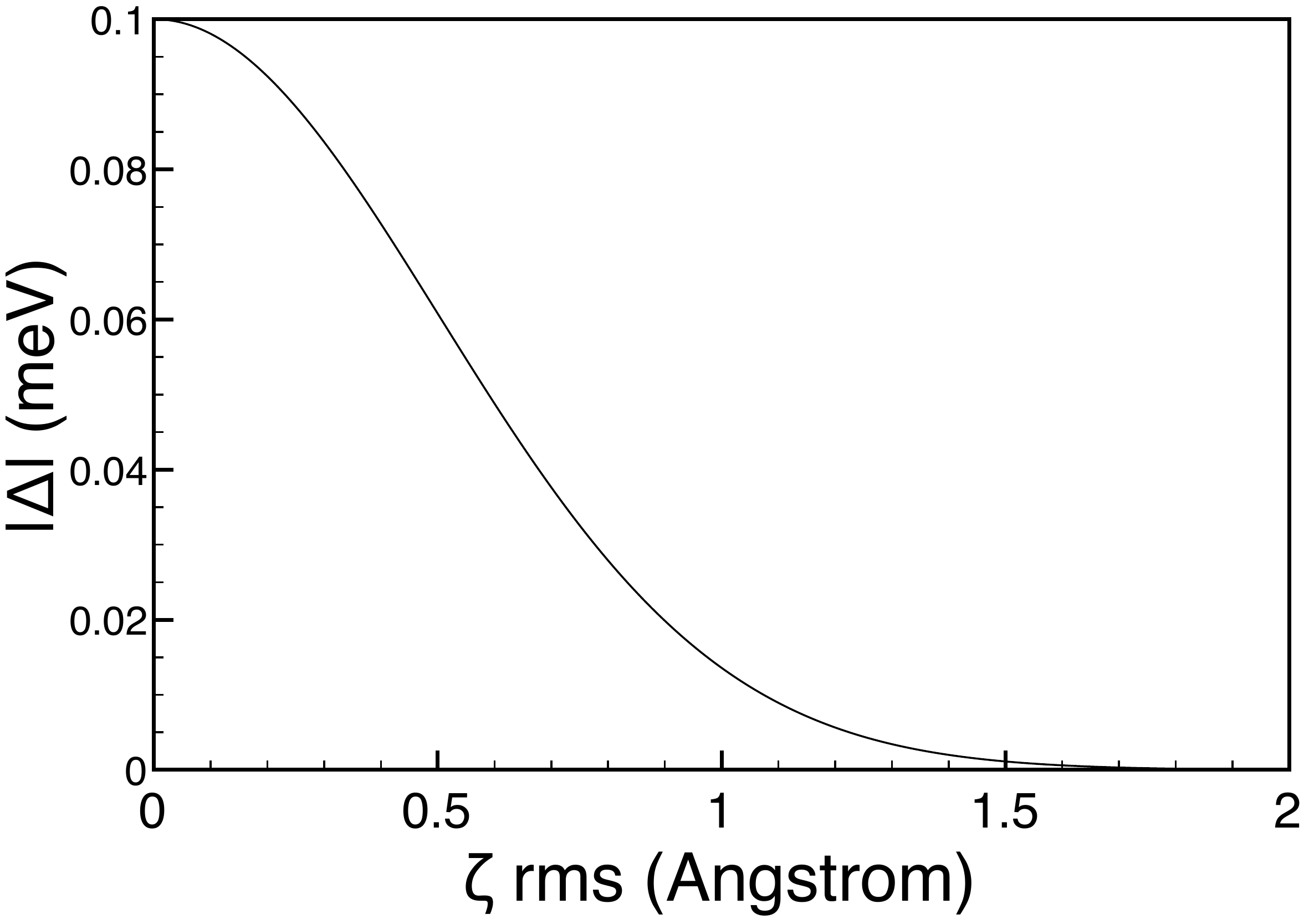}
\caption{Valley spliting $|\Delta|$ as a function of the rms fluctuations in the location of the interface $\zeta_{rms} \equiv \sqrt{\bkt{\zeta^2}}$. It is assumed that $\zeta$ is described by a Gaussian correlation function that varies on spatial scales considerably smaller than the extent of the quantum dot. An optimal value of $\Delta$ of 0.1meV has been assumed, as determined in Sec. \ref{sec:VOC} for a sharp flat interface.}
\label{Delta_vs_zeta_rms}
\end{figure}

If the roughness varies over a length scale of a few $\AA$, then the average of $e^{2i k_0 \, \zeta}$ over a quantum dot is equivalent to an average over the $xy$-plane, $\bkt{\zeta}_{QD} = \bkt{\zeta}$, which in turn is equivalent to an average over the realizations of $\zeta$, that is, it is the characteristic function of $\zeta$. If one assumes $\zeta$ to be described by a Gaussian function, then the characteristic function of $\zeta$ is given by $e^{-\kappa^2\bkt{\zeta^2}}$, where $\bkt{\zeta^2}$ is the variance of $\zeta$. In this case, since 
\begin{equation}
\frac{1}{\sqrt{2\pi\bkt{\zeta^2}}}\int d\zeta e^{i \kappa \, \zeta}\, e^{-\frac{\zeta^2}{2\bkt{\zeta^2}}} = e^{-\frac{\kappa^2\bkt{\zeta^2}}{2}},
\end{equation}
which implies that $\Delta$ is reduced by a factor of $e^{-\frac{\kappa^2\bkt{\zeta^2}}{2}}$. This is plotted in Fig.~\ref{Delta_vs_zeta_rms}. The reduction of $\Delta$ is in qualitative agreement with the conclusions of Ref.\ \onlinecite{Srinivasan_APL08}, although the spirit of Ref.\ \onlinecite{Srinivasan_APL08} is different from this work and a direct comparison is not possible. Similarly, a smeared interface potential can mimic roughness, allowing the position of the interface to be spread out over a range of spatial locations, \cite{Saraiva_PRB09} and resulting in a decrease in $\Delta$.

The variance of $\zeta(x,y)$ can be extracted from experiment, and its magnitude is critical. For $\sqrt{\bkt{\zeta^2}} = 2\AA$, the valley splitting $\tilde{\Delta}$ is suppressed by a factor of $e^{-8} = 3\times 10^{-4}$. On the other hand, for roughness of the order of $\zeta_{rms} = 0.7 \AA$, the exponential factor is $e^{-0.98} \approx 0.38 $. The optimal value of $\tilde{\Delta} \approx 0.1$meV that we have found above for a sharp flat interface would be reduced to $\approx 40 \mu$eV, which should be visible in a dilution refrigerator. At the same time we recall that the numerical calculations of Ref.\ \onlinecite{Saraiva_PRB09} found an optimal $\tilde{\Delta}$ of $\approx$ 0.25meV, which would be reduced to 0.1meV when $\zeta_{rms} = 0.7 \AA$ is taken into account. We will use this value in our estimates below. 

Thus far the magnitude and phase of $\tilde{\Delta}$ are the same on each dot, so $\tilde{t}_{+-} = 0$ everywhere in this section while $\tilde{t}_{0++} = \tilde{t}_{0--} = \tilde{t}_0$ and $\tilde{t}_{++} = \tilde{t}_{--} = \tilde{t}$. The term $\bkt{V(x,y) \, e^{2 i \,k_0\, \zeta}}_{QD}$ in turn must be taken as a phenomenological parameter to be measured experimentally. Noting that $\tilde{v}^D_{zz} = \tilde{v}^D_{\bar{z}\bar{z}}$ we find
\begin{equation}
\begin{array}{rl}
\displaystyle \tilde{v}^D_{\pm\pm} = & \displaystyle \tilde{v}^D_{zz} \pm (1/2)\, (e^{i\phi_D}\tilde{v}^D_{z\bar{z}} + e^{-i\phi_D}\tilde{v}^D_{\bar{z}z}) \\ [3ex]
\displaystyle \tilde{v}^D_{+-} = & \displaystyle (1/2)\, (e^{-i\phi_D}\tilde{v}^D_{\bar{z}z} - e^{i\phi_D}\tilde{v}^D_{z\bar{z}}).
\end{array} 
\end{equation}
If $\Delta$ is approximately imaginary so that $\phi \approx \pi/2$ these expressions simplify to 
\begin{equation}\label{vsimplified}
\begin{array}{rl}
\displaystyle \tilde{v}^D_{\pm\pm} = & \displaystyle \tilde{v}^D_{zz} \mp {\rm Im} \, \tilde{v}^D_{z\bar{z}} \\ [3ex]
\displaystyle \tilde{v}^D_{+-} = & \displaystyle -(i/2)\, {\rm Re}\, \tilde{v}^D_{z\bar{z}}.
\end{array} 
\end{equation}
It is seen that $v^D_{\pm\pm}$, and therefore $\tilde{v}_{zz}$, gives the energy offset between two dots that otherwise have the same valley splitting. Knowledge of $|\tilde{v}_{++}|$, which can be obtained from experiment, allows this diagonal term, important in inducing transitions between valley eigenstates, to be estimated. To summarize, for short interface correlations $\zeta$ gives us the overall magnitude of $\tilde{\Delta}$, while $v$ gives us the difference in $\tilde{\Delta}$ between dots. Below we consider the case when the local roughness matrix elements $v$ are much smaller than $\Delta$. 

\subsection{One-electron transitions}
\label{sec:1e2d} 

We study first the case of one electron on two dots. The physics depends on the relative magnitude of $\tilde{v}$ and $\tilde{t}_{0}$. For $\tilde{t}_{0} \gg \tilde{v}$, we diagonalize the Hamiltonian $H_{1e2d}$ with all the roughness terms set to zero, and denote the eigenenergies by $\tilde{\lambda}^{<}_\pm$, $\tilde{\lambda}^{>}_\pm$. In this scenario, the lowest energy state, which has energy $\tilde{\lambda}^{<}_-$, which is the state most easily initialized, does not cross any other state. Its closest approach to another state is of the order of $\approx \tilde{t}_0$. Since $\tilde{t}_0\gg \tilde{v}$ by assumption, valley manipulation by means of interface roughness is not feasible in this setup.

For $\tilde{v} \gg \tilde{t}_0$, the \textit{bare} Hamiltonian is now $H_{1e2d}$ with the local roughness terms but without interdot tunneling, in other words $\tilde{t}_0$ initially set to zero. The eigenstates of the bare Hamiltonian are localized on the left or the right dot. The eigenenergies are denoted $\tilde{\lambda}^{D}_<$, $\tilde{\lambda}^{D}_>$. Since $\tilde{t}_0$ is set to zero in the bare Hamiltonian, it is trivial to show that the eigenstates of $H_{1e2d}$ in this case do not depend on $\delta$. It is assumed that one initializes the state with energy $\tilde{\lambda}^R_<$ and sweeps the detuning until one reaches the point where this level crosses the state with energy $\tilde{\lambda}^L_<$. At this point the tunneling matrix element between these two levels is
\begin{equation}
\tilde{t}^{LR}_{<<} = \frac{\tilde{t}_0 \, (\tilde{v}_L^* \tilde{v}_R + \tilde{\lambda}^L_<\tilde{\lambda}^R_<)}{\sqrt{\tilde{\lambda}^{L2}_< + |\tilde{v}_L|^2}\sqrt{\tilde{\lambda}^{R2}_< + |\tilde{v}_R|^2}}.
\end{equation}
The critical crossing occurs at the point where $\lambda^R_< $, which depends linearly on the detuning, equals $\lambda^L_<$, which does not depend on the detuning. This condition yields a value of the detuning of
\begin{equation}
\begin{array}{rl}
\displaystyle \delta = & \displaystyle \frac{\tilde{\Delta}^L_+ + \tilde{\Delta}^L_-  - \tilde{\Delta}^R_+ - \tilde{\Delta}^R_- + \sqrt{(\tilde{\Delta}^L_+ + \tilde{\Delta}^L_-)^2 + 4|\tilde{v}^L_{+-}|^2} }{2} \\ [3ex]
- & \displaystyle \frac{\sqrt{(\tilde{\Delta}^R_+ + \tilde{\Delta}^R_-)^2 + 4|\tilde{v}^R_{+-}|^2}}{2}.
\end{array}
\end{equation}
The detuning is kept at this value for an amount of time $\approx \hbar/\tilde{t}^{LR}_{<<}$. The state of the electron after this time can be probed by sweeping the detuning back to $\delta \gg 0$, then doing charge sensing using a quantum point contact. This will reveal whether the electron has tunneled onto the left dot or remained on the right dot.

For weak local roughness the situation described here is simply ordinary single-particle tunneling including a roughness correction to the energy eigenvalues. The eigenstates depend on $\tilde{v}$, which is not known a priori. Yet the time scale governing the state mixing is given by the single-particle tunneling matrix element $\tilde{t}_0$, which is generally small. The size of this matrix element can be further reduced by the barrier gate, thus enabling experiment to increase the time scale for operations to a value which can be reliably handled in the laboratory.

The above demonstrates that, although intervalley transitions are extremely slow on one dot, they can be induced by interface roughness in a coherent manner. Nevertheless, operations using the single-electron two-dot spectrum suffer from the ambiguous initialization problems that hamper attempts to manipulate one-electron states. Initialization of $\tilde{D}_-$ is ambiguous since $\tilde{\Delta}_D$ may not be known. Moreover, one disadvantage of $\tilde{t}_0$ being small is that it is difficult to know the relative magnitude of $\tilde{t}_0$ and $\tilde{v}$ a priori.

\subsection{Two-electron transitions}
\label{sec:2e2d} 

For strong tunneling we first diagonalize the Hamiltonian with all the roughness parameters set to zero, $\tilde{v}_{\pm\pm} = \tilde{v}_{\pm\mp} = \tilde{t}_{\pm\mp} = 0$. We obtain in this way the basis of eigenstates for the strong tunneling case, which coincides with the singlet and triplet states used in Ref.\ \onlinecite{Culcer_10}. We work in this basis and add the roughness terms $\tilde{v}_{\pm\pm}$ and $\tilde{v}_{\pm\mp}$ perturbatively. Since interface roughness does not mix singlets with triplets, we will study the singlets and the triplets separately. Since the valley-orbit coupling is the same on both dots, we still have $\tilde{t}_{\pm\mp} = 0$. All relevant quantities are defined in Appendix \ref{app:eigenstates}.

For strong roughness one would have to diagonalize the Hamiltonian with all tunneling parameters set to zero, obtaining the basis of strong roughness eigenstates. One would then add the intravalley tunneling terms $\tilde{t}_{\pm\pm}$ as a perturbation. Theoretically, the possibility also exists that on one dot $\tilde{v} \gg \tilde{t}$, while on the other $\tilde{v} \ll \tilde{t}$. Yet we have noted that the tunneling matrix element $\tilde{t}$ contains a contribution due to the Coulomb interaction, denoted by $\tilde{s}$, which greatly enhances its magnitude with respect to $\tilde{t}_0$. One can obtain insight from recent noise measurements on Si QDs, \cite{Liu_Private} in which dangling bonds give rise to fluctuations in the confinement and barrier potentials. Such noise measurements give an indication of the magnitude of the potential fluctuations one can expect from dangling bonds, as well as random impurities, which both contribute to $\tilde{v}$. The size of the fluctuations has been measured to be  $\approx 0.45 \mu$eV,\cite{Liu_Private} whereas we expect $\tilde{t}$ to be of the order of tens of $\mu$eV. \cite{Culcer_10} The case $\tilde{v} \gg \tilde{t}$ is therefore much less likely in this setup and we do not consider it in what follows. 

\begin{figure}[tbp]
\includegraphics[width=\columnwidth]{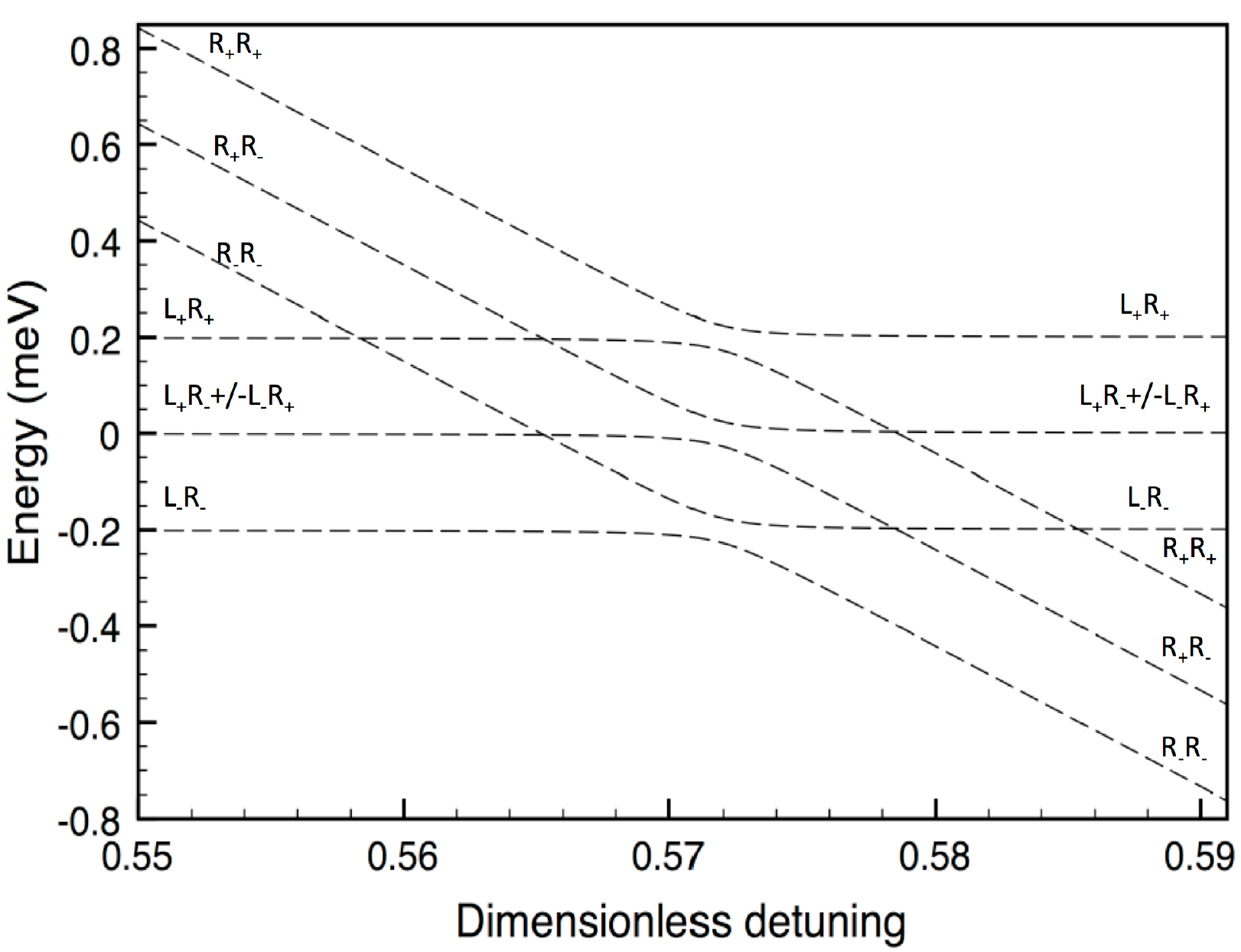}
\caption{Singlet levels in a double quantum dot with a valley splitting of 0.1meV in the absence of interface roughness, as determined in Ref.\ \onlinecite{Culcer_PRB09}.}
\label{Singlets}
\end{figure}

\begin{figure}[tbp]
\includegraphics[width=\columnwidth]{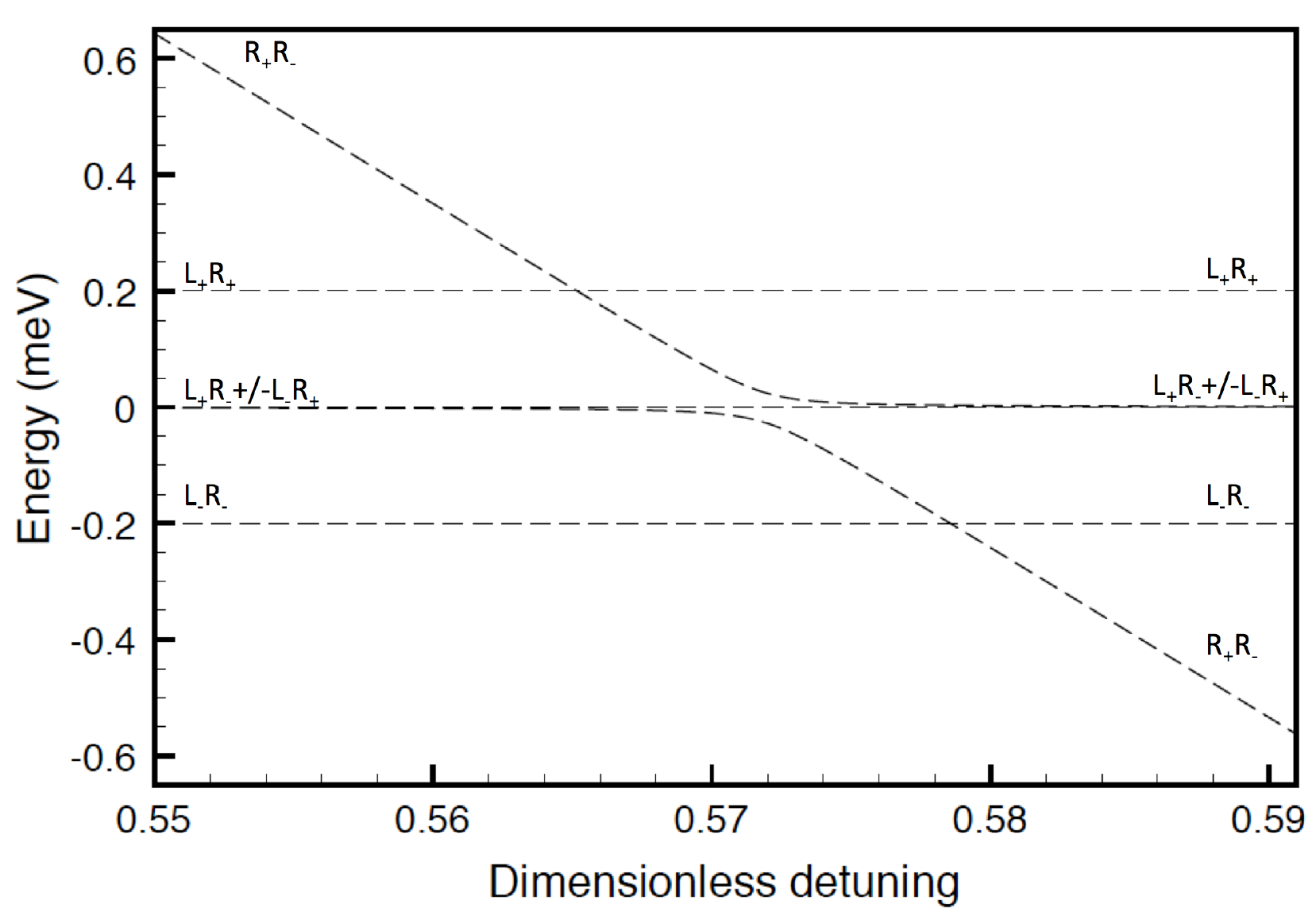}
\caption{Triplet levels in a double quantum dot with a valley splitting of 0.1meV in the absence of interface roughness, as determined in Ref.\ \onlinecite{Culcer_PRB09}.}
\label{Triplets}
\end{figure}

Since roughness is spin-independent and does not mix singlets with triplets, we discuss the singlet and triplet states separately. The bare singlet and triplet energy levels in the absence of roughness have been calculated in Ref.\ \onlinecite{Culcer_PRB09}. For convenience we have re-plotted them in this paper, showing the singlet and triplet levels separately in Figs.\ \ref{Singlets} and \ref{Triplets}. To determine which levels can be mixed by roughness we need to identify pairs of levels that cross and find out whether a nonzero matrix element of interface roughness can couple them. The singlet energy levels for $\tilde{t} \gg \tilde{v}$ are plotted in Fig.\ \ref{Singlets}. Here it is more convenient to use the dimensionless detuning as $\delta/(2d\tilde{\varepsilon}_0)$, where $d = X_0/a$ is the dimensionless half-interdot-distance. As Fig.\ \ref{Singlets} shows, the lowest energy singlet levels do not cross. Even in the case of large $\tilde{\Delta}$, when the lowest singlet state is easily initialized, there is no convenient point in the singlet energy level spectrum at which one can controllably induce mixing between valley eigenstates with different valley composition. The singlet states for $\tilde{t} \gg \tilde{v}$ are not suitable for valley manipulation by means of interface roughness.

We investigate next the triplet states for $\tilde{t} \gg \tilde{v}$. It is easiest to access the polarized triplets $\tilde{T}^+$ by starting out on one dot (for example $R$) and applying a large magnetic field to lower the energy of the $\tilde{T}^+$ states. This results in unambiguous initialization, regardless of the magnitude of the valley splitting, and is easily accomplished experimentally since $g=2$ in Si. Under these circumstances one needs to consider only the manifold of $\tilde{T}^+$ states, which comprises five wave functions. The triplet energies are plotted Fig. \ref{Triplets}. Far away from the anticrossing (at large detuning $\tilde{\delta} \gg 0 $) the eigenstates are still $\{\tilde{T}^{LR}_{- -}, \tilde{T}^{LR}_{++}, \tilde{T}^{LR}_{+-}, \tilde{T}^{LR}_{-+}, \tilde{T}^{RR}_{+-}\}$. As the detuning is swept towards $\tilde{\delta} \ll 0 $, the eigenstates evolve into combinations of these wave functions, which are given in the appendix. Out of the five levels of interest to us, $\tilde{T}^{RR}_{+-}$ evolves into $\tilde{T}^{<}_{+-}$, while $\tilde{T}^{LR}_{+-}$ remains unchanged. These two energy levels, $\tilde{T}^{<}_{+-}$ and $\tilde{T}^{LR}_{--}$ cross at a given value of the detuning. The critical crossing occurs at the point when the energy of $\tilde{T}^<_{+-}$, which depends on $\tilde{\delta}$, matches that of $\tilde{T}^{LR}_{--}$, which does not depend on $\tilde{\delta}$. Using the energies given in the appendix we obtain for the detuning at this crossing point $\tilde{\delta}  = (2\tilde{\Delta}^2 - \tilde{t}^2)/\tilde{\Delta}$. The matrix element mixing the two states is
\begin{equation}
\begin{array}{rl}
\displaystyle \tbkt{\tilde{T}^{LR}_{--}}{V}{\tilde{T}^<_{+-}} = & \displaystyle \frac{\tilde{t}}{\sqrt{\epsilon^{<2}_0 + 2\tilde{t}^2}} \, \bigg( \tilde{v}^L_{-+} - \tilde{v}^R_{-+} \bigg).
\end{array}
\end{equation}
Therefore, if one initializes $\tilde{T}^{RR}_{+-}$ and drives the system to this crossing point, the two levels can be mixed by interface roughness. The triplet $\tilde{T}^{RR}_{+-}$ is mixed by a small amount with $\tilde{T}^{LR}_{+-}$ and $\tilde{T}^{LR}_{-+}$, but this amount is negligible, since it is determined by the ratio $|\tilde{t}/\tilde{\delta}| \ll 1$. Therefore at the point where the two lowest triplet energy levels cross these levels are to a very good approximation $\tilde{T}^{RR}_{+-}$ and $\tilde{T}^{LR}_{--}$. The resulting configuration can be probed by charge sensing, since $\tilde{T}^+_{+-}$ goes back to (0,2) whereas $\tilde{T}^+_{--}$ remains in the (1,1) configuration. This process represents a coherent rotation of valley eigenstates. 

We note that for \textit{rapid} adiabatic passage (i.e. fast compared to $\hbar/\tbkt{\tilde{T}^{LR}_{--}}{V}{\tilde{T}^<_{+-}}$) the effect of roughness is analogous to a renormalization of the spectrum, which remains qualitatively the same, and valley eigenstates are not mixed. However for slow adiabatic passage valley eigenstates can be mixed. In effect one can use interface roughness to manipulate valleys in the same way the inhomogeneous nuclear field was used to manipulate spin singlet and triplet states.\cite{Petta_Science05} 

\section{Long interface correlations}
\label{sec:long}

This case of long interface correlations comprises variations in $\zeta$ on spatial scales comparable to that of the quantum dot. Recent experiments have indicated that this may be possible.\cite{Yoshinobu_JJAP94, Gotoh_APL02} In order to understand the implications of this possibility some preliminary considerations must be brought forth. Primarily among these, it is necessary to recall that, if both dots have the same valley-orbit coupling, they have the same valley splitting and the same valley eigenstates. Under such circumstances when electrons tunnel between the dots they tunnel between like valley eigenstates. Since this valley-orbit coupling can vary due to interface roughness, we should in principle expect both the amplitude and the phase of the valley-orbit coupling to be different between the two dots. Since the tunneling matrix element $\tilde{t}_0$ is defined as $\tilde{t}_0 = \tbkt{\tilde{L}}{H_0}{\tilde{R}}$, writing $\tilde{\Delta}_D = |\tilde{\Delta}_D|e^{-i\tilde{\phi}_D}$ gives $\tilde{t}_{0\pm\pm} = \tilde{t}_0[1 + e^{i(\tilde{\phi}_L - \tilde{\phi}_R)}]$ and $\tilde{t}_{0+-} = \tilde{t}_0[1 - e^{i(\tilde{\phi}_L - \tilde{\phi}_R)}]$. Since $\tilde{\Delta}_D$ does not depend explicitly on position, for \textit{2e2d}, using $\tilde{t}$ defined in Sec. \ref{sec:2eHam}, one has that $\tilde{t}_{\pm\pm} = \tilde{t}[1 + e^{i(\tilde{\phi}_L - \tilde{\phi}_R)}]$ and $\tilde{t}_{+-} = \tilde{t}[1 - e^{i(\tilde{\phi}_L - \tilde{\phi}_R)}]$. The matrix elements of the local roughness are unaltered by the relative location of the two dots. In all situations discussed in this paper, in the case $\tilde{v} \gg \tilde{t}$, whether the tunneling is intervalley or intravalley is a secondary consideration, since tunneling is still a perturbation compared to $\tilde{v}$. The key physics is provided by the matrix elements of $\tilde{v}$, which are the same regardless of whether the tunneling is intervalley or intravalley. Therefore the peculiar physics of the case in which interdot intervalley tunneling is comparable to interdot intravalley tunneling is manifest when $\tilde{t} \gg \tilde{v}$, and we focus on this regime for the remainder of this section. We will concentrate on the more enlightening \textit{2e2d} problem, in which the most striking consequences of interdot intervalley tunneling can be observed. In this section, since $\tilde{t}_{+-} \ne 0$, in the condition $\tilde{t} \gg \tilde{v}$ it is understood that $\tilde{t}$ represents the greater of $\tilde{t}_{--}$ and $\tilde{t}_{+-}$.

\begin{figure}[tbp]
\includegraphics[width=\columnwidth]{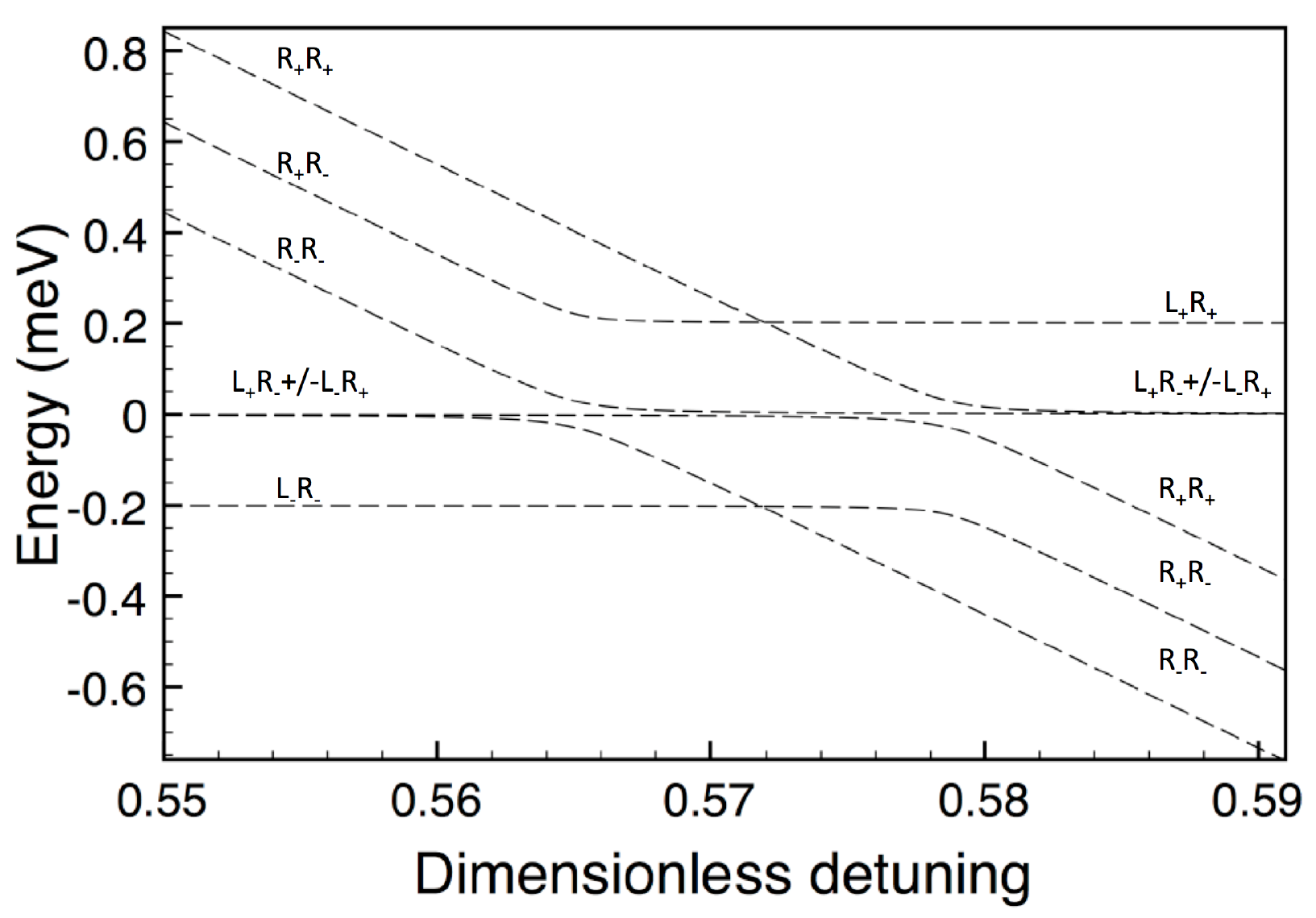}
\caption{Singlet states when $\tilde{\Delta}=$0.1meV, $\tilde{t}_{--} = 0$ and all the interdot tunneling is intervalley. On the far right side of the graph we recover the one-dot singlets, as in Fig.\ \ref{Singlets}.}
\label{Singlets_Intervalley}
\end{figure}

\begin{figure}[tbp]
\includegraphics[width=\columnwidth]{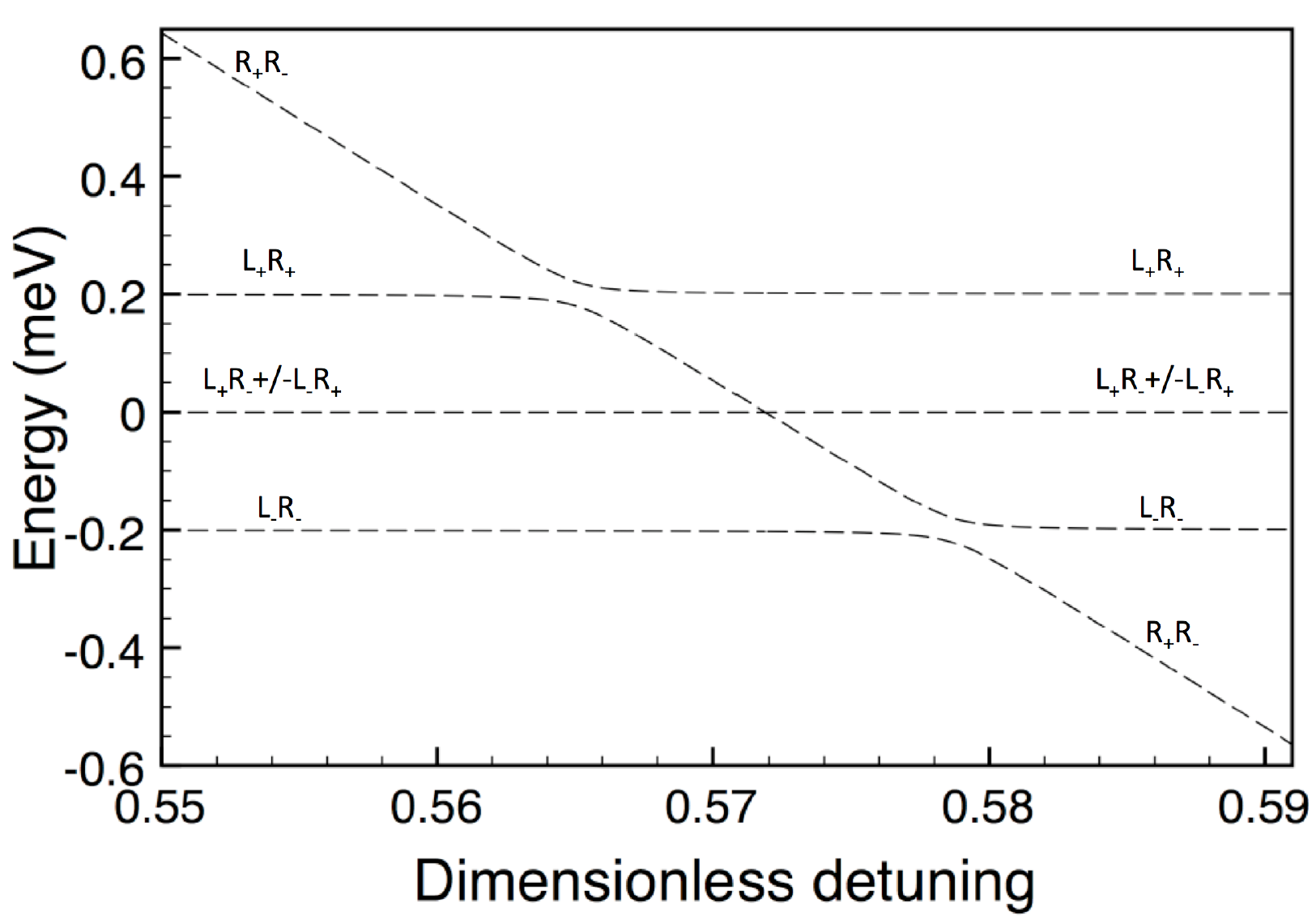}
\caption{Triplet states when $\tilde{\Delta}=$0.1meV, $\tilde{t}_{--} = 0$ and all the interdot tunneling is intervalley. On the far right side of the graph we recover the one-dot triplets, as in Fig.\ \ref{Triplets}.}
\label{Triplets_Intervalley}
\end{figure}

There are three qualitatively different situations to consider. In the easiest case, correlations stretch beyond the size of the entire DQD. The situation is almost identical to that of short interface correlations, except the value of $\tilde{\Delta}$ is not reduced from its \textit{ideal} value. If correlations occur on a spatial scale approximately equal to that of one dot, the magnitude of $\tilde{\Delta}$ is essentially the same on the two dots but its \textit{phase} is different: $\tilde{\phi}_L \ne \tilde{\phi}_R$, therefore $(\tilde{\phi}_L - \tilde{\phi}_R) \ne 0$. From the general definitions of $\tilde{t}_{\pm\pm}$ and $\tilde{t}_{+-}$ given above, it follows that in general both $\tilde{t}_{--}$ and $\tilde{t}_{+-}$ are nonzero, and their ratio will depend on the relative positions of the two dots. Since the spatial scale set by $2k_0$ is approximately 0.5$\AA$, this ratio can vary drastically from one sample to another, and in any one sample it is in effect a random number, which only experiment can determine reliably. This case cannot be treated perturbatively under any circumstances. We single out two important possibilities: (a) $\tilde{t}_{--} = 0$ and (b) $\tilde{t}_{--} = \tilde{t}_{+-}$. In case (a) all the interdot tunneling is intervalley tunneling. The situation is qualitatively completely different from the case when all the tunneling is intravalley, as Figs. \ref{Singlets_Intervalley} and \ref{Triplets_Intervalley} illustrate. None of the singlet levels cross other singlets, and none of the triplet levels cross other triplets. One cannot accomplish valley manipulation with either the singlets or the triplets. Case (b) is similar to (a) for the purposes of our paper. As long as the intervalley tunneling parameter is appreciable, the bottom two triplets are split. In this intermediate case the magnitude of $\tilde{t}_{--}$ and $\tilde{t}_{+-}$ is the same, meaning that the splitting of the bottom two states with different valley composition is the same as the splitting at the usual anticrossing given by $\tilde{t}_{--}$ (i.e. the anticrossing in Figs.\ref{Singlets} and \ref{Triplets}.) 

\begin{figure}[tbp]
\includegraphics[width=\columnwidth]{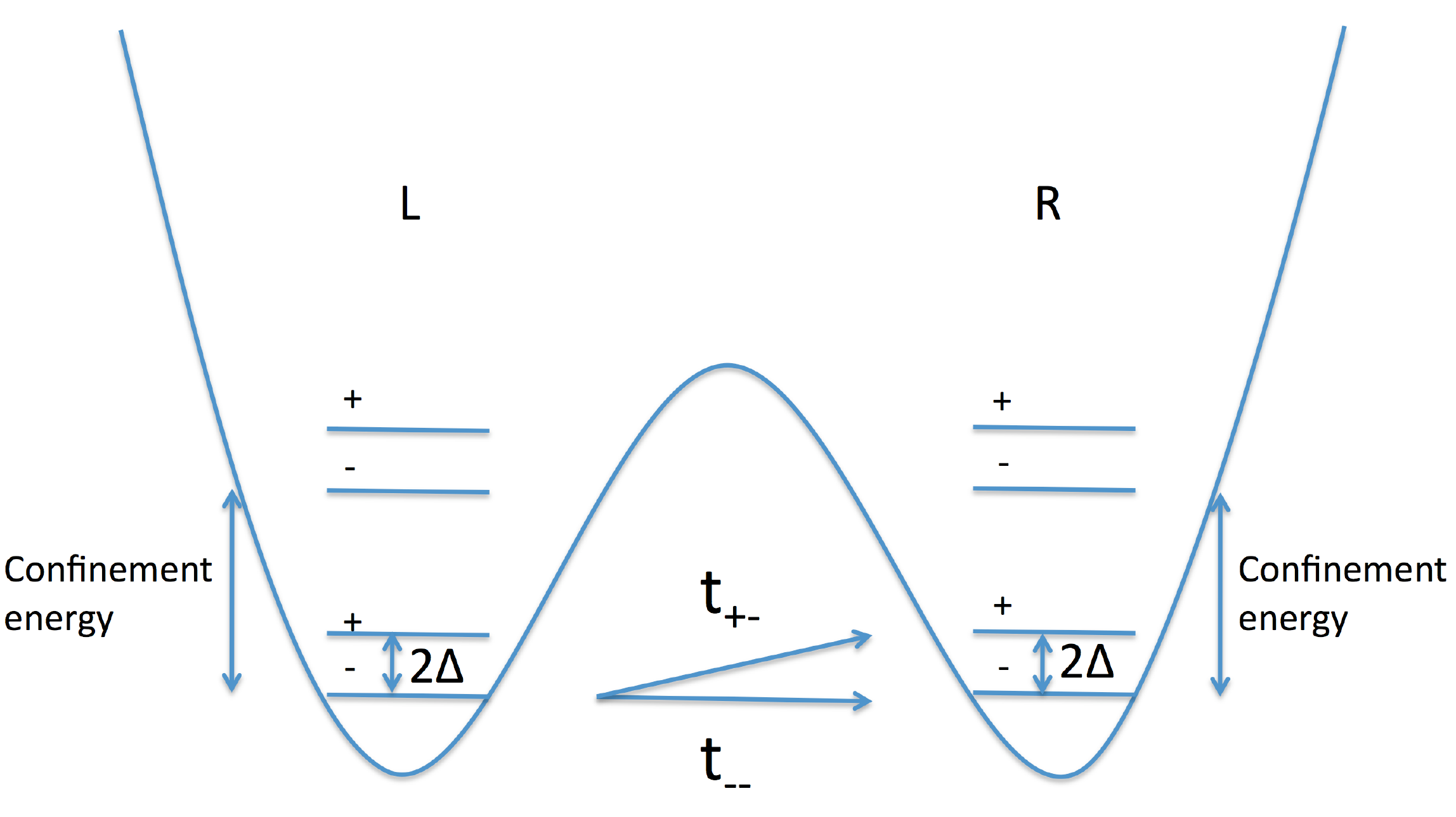}
\caption{Resonant tunneling in valley split quantum dots. The matrix elements for tunneling between like valley eigenstates and between opposite valley eigenstates are expected to be very different in magnitude. Resonant tunneling can detect these tunneling rates and give an accurate indication of their relative strength, allowing one to infer the length scale of interface roughness correlations.}
\label{RsnTnl}
\end{figure}

The analysis presented underscores the importance of determining the relative value of $\tilde{t}_{--}$ and $\tilde{t}_{+-}$. This can be accomplished by setting up the DQD for a resonant tunneling experiment, as in Fig. \ref{RsnTnl}. It is necessary first to do a transport experiment through each individual dot in order to identify the single-dot energy levels. In an ideal situation $k_BT \ll \tilde{\Delta} \ll \tilde{\varepsilon}_D$, and the levels on each dot have the form shown schematically in Fig. \ref{RsnTnl}. Subsequently, one can do a resonant tunneling experiment through the double quantum dot. The first peak in the resonant tunneling current will be between the lowest two levels and will give $\tilde{t}_{--}$. The levels of the left dot are subsequently held fixed while the levels of the right dot are scanned. The next peak in the resonant tunneling current should then give $\tilde{t}_{+-}$. In addition to providing values for the tunneling parameters, such an experiment should also give a good indication of the length scale characterizing interface roughness correlations. If $\tilde{t}_{+-}$ is close to zero it indicates that the correlation length of interface roughness is either much smaller than the size of a quantum dot or much larger than it. On the other hand, if $\tilde{t}_{+-}$ is measurable and of comparable magnitude to $\tilde{t}_{--}$ it indicates that the phase of $\tilde{\Delta}$ is different on the two dots and, from our discussion, roughness correlations occur on a scale comparable to that of a quantum dot. For this experiment, as for the valley manipulation scheme proposed in this paper, it is necessary to know the value of the valley splitting so that valley-split levels can be unambiguously identified. Two potential risks are associated with such a resonant tunneling experiment. Firstly, if the valley splitting is smaller than the typical broadening of the current peaks, then the signals due to $\tilde{t}_{--}$ and $\tilde{t}_{+-}$ may merge into one peak. In this case, the current peaks can be expected to have the same magnitude between any pair of levels on the left and right dots. Secondly, it may happen that one sample will have $|\tilde{t}_{--}| = |\tilde{t}_{+-}|$ within experimental error. Aside from the fact that such a risk is small, it can be overcome by performing the experiment on several samples. Furthermore, since $2\tilde{\Delta}$ is expected to be smaller than the confinement energy, the two tunneling parameters can be unambiguously identified.

\section{Discussion}
\label{sec:discussion}

\subsection{Overview of length scale considerations}

Our analysis makes evident the fact that the length scale characterizing interface roughness correlations plays an important role in determining the effect of roughness on the valley-orbit coupling and intervalley dynamics of double quantum dots. The two possible scenarios, interface correlations of the order of the lattice constant and greatly exceeding the lattice constant, produce vastly different physics on a qualitative level, as well as quantitatively different results. 

When the length scale of interface roughness correlations is of the order of a few lattice constants the magnitude of the valley-orbit coupling differs slightly on the two dots due to the local roughness, yet its phase is the same. All the interdot tunneling occurs between like valley eigenstates, that is, $+$ to $+$ and $-$ to $-$, while interdot intervalley tunneling is suppressed. We distinguish between two limiting physical situations depending on the strength of the local roughness relative to the interdot tunneling matrix element. Weak local roughness is characterized by $\tilde{t}_0 \gg \tilde{v}$ for \textit{1e2d} and $\tilde{t} \gg \tilde{v}$ for \textit{2e2d}. In the \textit{1e2d} setup the lowest energy levels do not cross, they come to within a distance $\approx \tilde{t}_0$ of each other. Since $\tilde{t}_0 \gg \tilde{v}$, the local roughness cannot mix these levels.

A feasible valley manipulation scheme using the singlet states cannot be devised either. Firstly, the singlets are only separated clearly from other levels in the (0,2) configuration. Secondly, there is no crossing of singlet levels that could be exploited. At the points at which singlets cross other triplet states are also close by. Even though roughness does not mix singlets and triplets, other mechanisms, such as the hyperfine coupling to the nuclei, could. On the other hand, the lowest \textit{2e2d} triplets do cross, providing a scheme that is experimentally accessible. By applying a magnetic field experiment can unambiguously access the $\tilde{T}^+$ triplet corresponding to the setup in which both electrons are on one dot. Once this state is initialized the detuning between the dots is ramped up until this state crosses the next-lowest energy state. At this point the local roughness can mediate transitions between these two levels, and sets the time scale for this mixing, which will be estimated below. 

For strong local roughness, $\tilde{v} \gg \tilde{t}$, it is obvious that the lowest energy levels cross in all setups. There is a tunneling matrix element between the two lowest energy levels, which in all three setups have different valley composition. The time scale for this level mixing is set by the interdot tunneling, in contrast to the weak local roughness case, in which it was set by the local roughness parameter $\tilde{v}$. On the one hand tunneling can be adjusted by means of the gate that controls the interdot barrier, allowing some experimental control.

When the length scale of interface roughness correlations is of the order of the size of a quantum dot, the magnitude of the valley-orbit coupling will again be slightly different between the two dots, yet its phase can in principle also be different. The relative phase of the valley-orbit coupling is determined by the value of the function $e^{2ik_0\zeta}$ on each dot and, where $2k_0$ sets a length scale of $\approx 0.5\AA$. The relative phase of the valley orbit coupling determines the relative magnitude of interdot tunneling between like valley eigenstates and interdot tunneling between opposite valley eigenstates. Given the random nature of $\zeta$, the extremely short length scale set by $2k_0$ and the impossibility of achieving control over such short distances in the foreseeable future, the relative strengths of tunneling between like and opposite valley eigenstates under these circumstances is in effect a random number, expected to vary greatly from one sample to another. 

\subsection{Strong intervalley coupling limit}

The extreme cases considered, $\tilde{t}_{+-} = \tilde{t}_{--}$ and $\tilde{t}_{--} = 0$ are rather similar qualitatively. Once again one must consider how the strength of the local roughness compares to the interdot tunneling. For $\tilde{t} \gg \tilde{v}$ the lowest energy levels do not cross in any of the three possible setups. If $\tilde{v} \gg \tilde{t}$ the bare Hamiltonian is the same as when the tunneling is predominantly intravalley. We have established that the lowest energy levels do cross in all setups, yet the tunneling between them is determined by the relative strength of  $\tilde{t}_{--}$ and $\tilde{t}_{+-}$. Since this is a random number any valley manipulation scheme under these circumstances must be regarded as overly ambitious. If interface roughness correlations occur on the scale of a quantum dot the valleys are mixed randomly and any attempt at coherent manipulation of valley eigenstates would be futile.

One important result of our study is the fact that variations in the position of the interface (given by $\bkt{\zeta^2}$) and variations in the local roughness (given by $\tilde{v}$) influence the valley-orbit coupling and interdot dynamics in vastly different ways. The length scale of the interface misalignment across a double quantum dot gives rise to two qualitative scenarios: short correlations of interface roughness, over the scale of a few lattice constants, and long correlations of interface roughness, over the scale of a quantum dot. For short correlations the valley-orbit coupling is suppressed by a factor $e^{-\frac{\kappa^2\bkt{\zeta^2}}{2}}$, with $\kappa$ determined by the locations of the valleys in reciprocal space, and $\bkt{\zeta^2}$ the variance of the fluctuations in the interface location. In the absence of local roughness, the valley-orbit coupling in a double quantum dot is the same on each dot. For long correlations the phase of the valley-orbit coupling can also be different between the two dots, such that interdot tunneling between opposite valley eigenstates is enabled. In this case the relative phase of the valley-orbit coupling cannot be controlled and the relative magnitude of interdot tunneling between like and opposite valley eigenstates is essentially a random variable. These findings are general and applicable to all systems where valley physics is important. 

\subsection{Ideal scenario}

The analysis presented in this work suggests that the ideal scenario for inducing transitions between valley eigenstates is to use the polarized triplet $\tilde{T}^+$ states in a DQD in which $\tilde{t} \gg \tilde{v}$. To determine the time scale for these transitions we need to estimate the magnitude of $\tilde{v}$, or more precisely its variation across a DQD (which may be due to dangling bonds or impurities.) In the absence of experimental data, it is necessary to estimate the variation in $\tilde{\Delta}_D$ between the two dots. We assume $\tilde{\Delta}_D = 0.1$ meV and, to cover a wider range of experimental possibilities, we consider variations in $\tilde{\Delta}_D$ of $1-10\%$. As shown above, knowledge of the difference in $\tilde{\Delta}_D$ enables one to determine the difference in $\tilde{v}_{zz}$ between the dots (we assume $\tilde{v}^D_{zz} \gg {\rm Im} \, \tilde{v}^D_{z\bar{z}}$ in Eq. \ \ref{vsimplified}, so that the difference in $\tilde{\Delta}_D$ is given by the difference in $\tilde{v}_{zz}$ only.) Moreover, we can use the evaluations of $\tilde{\Lambda}_D$ and $\tilde{\Delta}_D$ to infer a series of facts about the matrix elements of $V(x,y)$. The matrix elements $\tilde{v}_{zz}$ and $\tilde{v}_{z\bar{z}}$ follow the same pattern as $\tilde{\Lambda}_D$ and $\tilde{\Delta}_D$ respectively, except that instead of $U_0$ we have the average of $V(x,y)$ over a quantum dot. Therefore, based on the ratio of $\tilde{\Delta}_D$ and $\tilde{\Lambda}_D$, we conclude that the ratio of the off-diagonal matrix element $\tilde{v}_{z\bar{z}}$ to the diagonal matrix element $\tilde{v}_{zz}$ can range between $\approx 0.1 - 0.3$ depending on whether SiGe is used or SiO$_2$. We estimate in addition the real part of $\tilde{v}_{z\bar{z}}$ to be one order of magnitude smaller than the imaginary part. The roughness average reduces this by another order of magnitude, with a further reduction by an order of magnitude arising from the multiplicative factor $\tilde{t}/\tilde{\Delta}_D$. One ultimately expects $\tilde{v}$ in the range $ \approx 0.1 - 1$ neV, giving a time scale of approximately 1-10 $\mu$s. This figure represents the time scale for transitions from one valley eigenstate to another. Gating on such a time scale is easily accomplished experimentally.\cite{Petta_Science05}

\subsection{Important regions of the energy spectrum}

In all scenarios discussed in this work there are three relevant regions of the energy level spectrum as a function of interdot detuning $\tilde{\delta}$. These are the two extreme cases $\tilde{\delta} \ll 0$, $\tilde{\delta} \gg 0$, as well as a third value $\tilde{\delta}$ at which the lowest two energy levels cross, in the cases when such a crossing does occur. For \textit{1e2d} in the strong tunneling regime the two lowest energy levels as a function of detuning go through an avoided crossing where their minimum separation is of order $\tilde{t}$. The matrix element mixing these levels is of order $\tilde{v}$, and since $\tilde{v} \ll \tilde{t}$ the possibility of mixing these levels by means of interface roughness is remote. For \textit{2e2d} in the strong tunneling regime, the lowest singlet states also anticross and cannot be mixed, yet two triplet states with different valley compositions cross each other and can be mixed by roughness. In fact transitions between triplet states offer the best prospect of experimental manipulation of the valley degree of freedom using interface roughness. For strong roughness all the above scenarios (\textit{1e2d}, \textit{2e2d} singlet and triplet states) are very similar. Without interdot tunneling many energy levels cross, and in the neighborhood of the crossing point the lowest two energy levels can be mixed by $\tilde{t}$. Given that the tunneling parameter can be adjusted by means of the gate that controls the barrier between the dots, it can in principle be reduced to a time scale slow enough that a coherent valley experiment is possible with currently achievable gating times.

\section{Summary}

Interface roughness has two principal effects on quantum dots made in materials with multivalley energy spectra. Firstly, it leads to a suppression of the valley-orbit coupling by a factor $e^{-2k_0^2\bkt{\zeta^2}}$, where $\bkt{\zeta^2}$ is the variance of the fluctuations in the position of the interface. Secondly, in a double quantum dot it enables transitions between valley eigenstates. Such transitions occur in the dynamics of one and two electrons in a DQD. In the majority of processes discussed, one electron initialized into a particular valley eigenstate on one dot tunnels into the opposite valley eigenstate on the adjacent dot when the detuning between the dots is swept to a certain point of the energy spectrum and appropriate mixing time is allowed. The position of the mixing point and the value of the mixing time must be determined individually for each configuration -- \textit{1e1d}, \textit{2e2d} singlets and \textit{2e2d} triplets, and depends also on the relative magnitude of the local interface roughness and the relevant interdot tunneling matrix element. Based on our study of all possible configurations we conclude that the likeliest system to observe this phenomenon is provided by the triplet states in the \textit{2e2d} configuration, which may be reliably initialized and are estimated to mix on a time scale of 1-10 $\mu$s. It must be pointed out that, although the possibility of inducing transitions between valley eigenstates by means of interface roughness discussed in this article is one way of manipulating valleys, the scheme we have presented is not tunable. Instead it must be regarded as a first step in the attempt to achieve coherent control of valley eigenstates.

\begin{acknowledgements}

The authors would like to thank E. Kaxiras, Belita Koiller, A.~L.~Saraiva, Andr\'as P\'alyi, H.~W.~Jiang, M.~A.~ Eriksson, N.~M.~Zimmerman, Matthew House, Rogerio de Sousa, M.~S.~Carroll, M.~P.~Lilly, Mark Friesen, S.~N.~Coppersmith, Jason Kestner, L.~Cywi\'nski and Ted Thorbeck for stimulating discussions. This work is supported by LPS-NSA-CMTC. XH also acknowledges support by NSA/LPS via ARO.

\end{acknowledgements}

\appendix

\section{Valley-orbit coupling}
\label{app:VOC}

The main contribution to $\Delta_D$ comes from the interface potential. For a perfectly smooth and perfectly sharp interface this is
\begin{widetext} 
\begin{equation}
\begin{array}{rl}
\displaystyle \Delta_D = \tbkt{\Psi_z}{\mathcal{V}}{\Psi_{\bar{z}}} = & \displaystyle U_0 \, \int_{-\infty}^{\infty} \int_{-\infty}^{\infty} dx\, dy\, |\phi(x,y)|^2  \int_{-\infty}^{\infty} dz\, \Theta(-z)\, |\psi (z)|^2 \, e^{-2ik_z z}\, u^*_z ({\bm r}) \, u_{\bar{z}} ({\bm r}) \\ [3ex]
= & \displaystyle U_0 \, N^2 \, z_0^2 \sum_{{\bm K}{\bm Q}} \, c^{z*}_{{\bm K}} c^{\bar{z}}_{{\bm K} + {\bm Q}} \int_{-\infty}^{\infty} \int_{-\infty}^{\infty} dx\, dy\, |\phi(x,y)|^2 \,  e^{i{\bm Q}_\perp\cdot{\bm r}} \int_{-\infty}^0 dz \, e^{q_z z}.
\end{array}
\end{equation}
where $q_\xi = k_b + iQ_z - 2ik_\xi$. We can approximate the $x$ and $y$ integrals by $\delta$-functions of $Q_x$, $Q_y$, yielding 
\begin{equation}
\begin{array}{rl}
\displaystyle \tbkt{\Psi_z}{\mathcal{V}}{\Psi_{\bar{z}}} \approx & \displaystyle U_0 \, N^2 \, z_0^2 \sum_{{\bm K}, Q_z} \, \frac{c^{z*}_{{\bm K}} c^{-z}_{{\bm K} + Q_z{\bm z}}}{q_z}.
\end{array}
\end{equation}
The matrix element between the same valley eigenstates is given by
\begin{equation}
\begin{array}{rl}
\displaystyle \Lambda_D = \tbkt{\Psi_\xi}{\mathcal{V}}{\Psi_\xi} = & \displaystyle U_0 \, \int_{-\infty}^{\infty} \int_{-\infty}^{\infty} dx\, dy\, |\phi(x,y)|^2  \int_{-\infty}^{\infty} dz\, \Theta(-z)\, |\psi (z)|^2 \,  |u_\xi ({\bm r})|^2 \\ [3ex]
= & \displaystyle U_0 \, N^2 \, z_0^2 \sum_{{\bm K}{\bm Q}} \, c^{\xi*}_{{\bm K}} c^{\xi}_{{\bm K} + {\bm Q}} \int_{-\infty}^{\infty} \int_{-\infty}^{\infty} dx\, dy\, |\phi(x,y)|^2 \,  e^{i{\bm Q}_\perp\cdot{\bm r}} \int_{-\infty}^0 dz\, e^{(k_b + iQ_z)z}.
\end{array}
\end{equation}
In the presence of an interface $\psi (z) \rightarrow \psi (z - \zeta)$. In the presence of interface roughness 
\begin{equation}
\begin{array}{rl}
\displaystyle \Delta_D = & \displaystyle \int_{-\infty}^{\infty} \int_{-\infty}^{\infty} dx\, dy\, [U_0 + V(x,y)] \,  |\phi(x,y)|^2  \int_{-\infty}^{\infty} dz\, \Theta[-(z - \zeta)] \, |\psi (z - \zeta)|^2 \, e^{-2ik_z z}\, u^*_z ({\bm r}) \, u_{\bar{z}} ({\bm r}) \\ [3ex]
= & \displaystyle \sum_{{\bm K}{\bm Q}} \, c^{z*}_{{\bm K}} c^{\bar{z}}_{{\bm K} + {\bm Q}} \int_{-\infty}^{\infty} \int_{-\infty}^{\infty} dx\, dy\, [U_0 + V(x,y)] \, |\phi(x,y)|^2 \, e^{i{\bm Q}_\perp\cdot{\bm r}} \int_{-\infty}^{\infty} dz\, \Theta[-(z - \zeta)] \, |\psi (z - \zeta)|^2 \, e^{i(Q_z - 2k_z) z}.
\end{array}
\end{equation}
The $z$-integral is performed by first changing the variable of integration to $z' = z - \zeta$ (let also $\kappa_\xi = Q_z - 2k_\xi$) 
\begin{equation}
\begin{array}{rl}
\displaystyle \int_{-\infty}^{\infty} dz\, \Theta[-(z - \zeta)] \, |\psi (z - \zeta)|^2 \, e^{i \kappa_\xi z} = & \displaystyle \int_{-\infty}^{\infty} dz' \, \Theta(-z') \, |\psi (z')|^2 \, e^{i \kappa_\xi (z' + \zeta)} = e^{i \kappa_\xi \, \zeta}\int_{-\infty}^0 dz' \, |\psi (z')|^2 \, e^{i \kappa_\xi \, z'}.
\end{array}
\end{equation}
We always have $z' \le 0$ and $\psi (z') = N \, z_0 \, e^{\frac{\kappa_{ox} z'}{2}}$, so the integral is trivial
\begin{equation}
e^{i \kappa_\xi \, \zeta} \int_{-\infty}^0 dz' \, |\psi (z')|^2 \, e^{i \kappa_\xi \, z'} = \frac{N^2 \, z_0^2 \, e^{i \kappa_\xi \, \zeta}}{q_\xi}.
\end{equation}
\end{widetext}

\section{Matrix elements of roughness in the singlet and triplet manifolds}
\label{app:matrixel}

We will first give the matrix elements of the roughness potential between the \textit{bare} singlet and triplet states. These are in fact the same, so $\tbkt{\tilde{S}}{V}{\tilde{S}} = \tbkt{\tilde{T}}{V}{\tilde{T}}$ for all the cases below. There are no matrix elements of the roughness potential connecting singlet and triplet states. The random potential is written as $V ({\bm r}_1) + V ({\bm r}_2)$. The intrabranch matrix elements are
\begin{equation}
\begin{array}{rl}
\displaystyle \tbkt{\tilde{S}^{RR}_{++}}{V}{\tilde{S}^{RR}_{++}} = & \displaystyle 2 \tilde{v}^R_{++} \\ [3ex]
\displaystyle \tbkt{\tilde{S}^{RR}_{+-}}{V}{\tilde{S}^{RR}_{+-}} = & \displaystyle \tilde{v}^R_{++} + \tilde{v}^R_{--} \\ [3ex]
\displaystyle \tbkt{\tilde{S}^{LR}_{++}}{V}{\tilde{S}^{LR}_{++}} = & \displaystyle \tilde{v}^L_{++} + \tilde{v}^R_{++} \\ [3ex]
\displaystyle \tbkt{\tilde{S}^{LR}_{+-}}{V}{\tilde{S}^{LR}_{+-}} = & \displaystyle \tilde{v}^L_{++} + \tilde{v}^R_{--} \\ [3ex]
\displaystyle \tbkt{\tilde{S}^{LR}_{-+}}{V}{\tilde{S}^{LR}_{-+}} = & \displaystyle \tilde{v}^L_{--} + \tilde{v}^R_{++} \\ [3ex]
\displaystyle \tbkt{\tilde{S}^{LR}_{+-}}{V}{\tilde{S}^{LR}_{-+}} = & \displaystyle 0.
\end{array}
\end{equation}
The interbranch matrix elements are
\begin{equation}
\begin{array}{rl}
\displaystyle \tbkt{\tilde{S}^{RR}_{+-}}{V}{\tilde{S}^{RR}_{++}} = & \displaystyle \sqrt{2} \tilde{v}^R_{-+} \\ [3ex]
\displaystyle \tbkt{\tilde{S}^{RR}_{+-}}{V}{\tilde{S}^{RR}_{--}} = & \displaystyle \sqrt{2} \tilde{v}^R_{+-} \\ [3ex]
\displaystyle \tbkt{\tilde{S}^{LR}_{+-}}{V}{\tilde{S}^{LR}_{++}} = & \displaystyle \tilde{v}^R_{-+} \\ [3ex]
\displaystyle \tbkt{\tilde{S}^{LR}_{+-}}{V}{\tilde{S}^{LR}_{--}} = & \displaystyle \tilde{v}^L_{+-} \\ [3ex]
\displaystyle \tbkt{\tilde{S}^{LR}_{-+}}{V}{\tilde{S}^{LR}_{++}} = & \displaystyle \tilde{v}^L_{-+} \\ [3ex]
\displaystyle \tbkt{\tilde{S}^{LR}_{-+}}{V}{\tilde{S}^{LR}_{--}} = & \displaystyle \tilde{v}^R_{+-}. 
\end{array}
\end{equation}
All other matrix elements are assumed to be negligible.

\section{Singlet and triplet eigenstates of 2e2d for $\tilde{t} \gg \tilde{v}$}
\label{app:eigenstates}

We will use the following notation for the energy eigenstates
\begin{equation}
\begin{array}{rl}
\displaystyle \tilde{\varepsilon}^{>}_\pm = \frac{ -\tilde{\delta} + \sqrt{\tilde{\delta}^2 + 8\tilde{t}^2} }{2} \pm 2\tilde{\Delta}_0 \\ [3ex]
\displaystyle \tilde{\varepsilon}^{<}_\pm = \frac{ -\tilde{\delta} - \sqrt{\tilde{\delta}^2 + 8\tilde{t}^2} }{2} \pm 2\tilde{\Delta}_0 \\ [3ex]
\displaystyle \tilde{\varepsilon}^{>}_0 = \frac{ -\tilde{\delta} + \sqrt{\tilde{\delta}^2 + 8\tilde{t}^2} }{2} \\ [3ex]
\displaystyle \tilde{\varepsilon}^{<}_0 = \frac{ -\tilde{\delta} - \sqrt{\tilde{\delta}^2 + 8\tilde{t}^2} }{2}.
\end{array} 
\end{equation}
The singlet eigenstates are
\begin{equation}
\begin{array}{rl}
\displaystyle \tilde{S}^>_{\pm\pm} = & \displaystyle \frac{\tilde{\varepsilon}^>_\pm}{\sqrt{\tilde{\varepsilon}^{>2}_\pm + 2\tilde{t}^2}} \, \bigg( \frac{\tilde{t}\sqrt{2}}{\tilde{\varepsilon}^>_\pm} \, \tilde{S}^{LR}_{\pm\pm} + \tilde{S}^{RR}_{\pm\pm} \bigg) \\ [3ex]
\displaystyle \tilde{S}^<_{\pm\pm} = & \displaystyle \frac{\tilde{\varepsilon}^<_\pm}{\sqrt{\tilde{\varepsilon}^{<2}_\pm + 2\tilde{t}^2}} \, \bigg( \frac{\tilde{t}\sqrt{2}}{\tilde{\varepsilon}^<_\pm} \, \tilde{S}^{LR}_{\pm\pm} + \tilde{S}^{RR}_{\pm\pm} \bigg) \\ [3ex]
\displaystyle \tilde{S}^>_{+-} = & \displaystyle \frac{\tilde{\varepsilon}^>_0}{\sqrt{\tilde{\varepsilon}^{>2}_0 + 2\tilde{t}^2}} \, \bigg( \frac{\tilde{t}}{\tilde{\varepsilon}^>_0} \, \tilde{S}^{LR}_{+-} + \frac{\tilde{t}}{\tilde{\varepsilon}^>_0} \, \tilde{S}^{LR}_{-+} + \tilde{S}^{RR}_{+-} \bigg) \\ [3ex]
\displaystyle \tilde{S}^<_{+-} = & \displaystyle \frac{\tilde{\varepsilon}^<_0}{\sqrt{\tilde{\varepsilon}^{<2}_0 + 2\tilde{t}^2}} \, \bigg( \frac{\tilde{t}}{\tilde{\varepsilon}^<_0} \, \tilde{S}^{LR}_{+-} + \frac{\tilde{t}}{\tilde{\varepsilon}^<_0} \, \tilde{S}^{LR}_{-+} + \tilde{S}^{RR}_{+-} \bigg) \\ [3ex]
\displaystyle \tilde{S}^{anti}_{+-} = & \displaystyle \frac{1}{\sqrt{2}} \, \bigg( \tilde{S}^{LR}_{+-} - \tilde{S}^{LR}_{-+} \bigg).
\end{array}
\end{equation}
Two triplet eigenstates are $\tilde{T}^{LR}_{\pm\pm}$. The other three are
\begin{equation}
\begin{array}{rl}
\displaystyle \tilde{T}^>_{+-} = & \displaystyle \frac{\tilde{\varepsilon}^>_0}{\sqrt{\tilde{\varepsilon}^{>2}_0 + 2\tilde{t}^2}} \, \bigg( \frac{\tilde{t}}{\tilde{\varepsilon}^>_0} \, \tilde{T}^{LR}_{+-} - \frac{\tilde{t}}{\tilde{\varepsilon}^>_0} \, \tilde{T}^{LR}_{-+} + \tilde{T}^{RR}_{+-} \bigg) \\ [3ex]
\displaystyle \tilde{T}^<_{+-} = & \displaystyle \frac{\tilde{\varepsilon}^<_0}{\sqrt{\tilde{\varepsilon}^{<2}_0 + 2\tilde{t}^2}} \, \bigg( \frac{\tilde{t}}{\tilde{\varepsilon}^<_0} \, \tilde{T}^{LR}_{+-} - \frac{\tilde{t}}{\tilde{\varepsilon}^<_0} \, \tilde{T}^{LR}_{-+} + \tilde{T}^{RR}_{+-} \bigg) \\ [3ex]
\displaystyle \tilde{T}^{sym}_{+-} = & \displaystyle \frac{1}{\sqrt{2}} \, \bigg( \tilde{T}^{LR}_{+-} + \tilde{T}^{LR}_{-+} \bigg).
\end{array}
\end{equation}


\end{document}